\documentclass[aps, nofootinbib,superscriptaddress, preprintnumbers, longbibliography]{article}

\usepackage{tikz}
\usepackage{jcappub}
\usepackage[utf8]{inputenc}
\usepackage{amsmath,amsfonts,amssymb}
\usepackage{graphicx}
\usepackage[toc,page]{appendix}
\usepackage{cleveref}
\usepackage[loose]{units}
\usepackage{siunitx} 
\usepackage{booktabs}
\usepackage{mathtools}
\newcommand{\beqn}{\begin{equation}}
\newcommand{\eeqn}{\end{equation}}

\makeatletter
\gdef\@fpheader{}
\makeatother

\begin{document}

\title{
Lattice Simulations of Inflation}

\author[a,b]{Angelo Caravano,}
\author[c,d]{Eiichiro Komatsu,}
\author[e]{Kaloian D. Lozanov,}
\author[a,f]{Jochen Weller}

\affiliation[a]{
	Universit\"ats-Sternwarte M\"unchen, Fakult\"at f\"ur Physik, Ludwig-Maximilians Universit\"at,\\ Scheinerstr. 1, 81679 M\"unchen, Germany}
\affiliation[b]{Max-Planck-Institut f\"ur Physik (Werner-Heisenberg-Institut),\\
	F\"ohringer Ring 6, 80805 Munich, Germany}
\affiliation[c]{Max Planck Institute for Astrophysics, Karl Schwarzschild Str. 1, Garching, 85741, Germany}
\affiliation[d]{Kavli Institute for the Physics and Mathematics of the Universe (Kavli IPMU, WPI), University of Tokyo, Chiba 277-8582, Japan}
\affiliation[e]{Illinois Center for Advanced Studies of the Universe \& Department of Physics, University of Illinois at Urbana-Champaign, Urbana, IL 61801, USA.}

\affiliation[f]{Max Planck Institute for Extraterrestrial Physics,\\
	Giessenbachstr. 1, 85748 Garching, Germany}

\emailAdd{caravano@usm.lmu.de}

\abstract{The scalar field theory of cosmological inflation constitutes nowadays one of the preferred scenarios for the physics of the early universe. In this paper we aim at studying the inflationary universe making use of a numerical lattice simulation. Various lattice codes have been written in the last decades and have been extensively used for understating the reheating phase of the universe, but they have never been used to study the inflationary phase itself far from the end of inflation (i.e. about 50 e-folds before the end of inflation). In this paper we use a lattice simulation to reproduce the well-known results of some simple models of single-field inflation, particularly for the scalar field perturbation. The main model that we consider is the standard slow-roll inflation with an harmonic potential for the inflaton field. We explore the technical aspects that need to be accounted for in order to reproduce with precision the nearly scale invariant power spectrum of inflaton perturbations. We also consider the case of a step potential, and show that the simulation is able to correctly reproduce the oscillatory features in the power spectrum of this model. Even if a lattice simulation is not needed in these cases, that are well within the regime of validity of linear perturbation theory, this sets the basis to future work on using lattice simulations to study more complicated models of inflation. 
}

\maketitle

\section{Introduction}
 Inflation is the accelerated expansion of the early universe originally introduced as a solution to the horizon and flatness problems of Cosmology \cite{PhysRevD.23.347,Sato:1980yn,Linde:1981mu,PhysRevLett.48.1220,STAROBINSKY198099}. In the standard picture this early accelerated expansion is driven by a scalar field, the inflaton, which acts as a source with a negative pressure. The standard theory of inflation predicts scalar and tensor perturbations in the early universe as quantum vacuum fluctuations \cite{Starobinsky:1979ty,Mukhanov:1981xt,HAWKING1982295,PhysRevLett.49.1110,STAROBINSKY1982175,Abbott:1984fp}. While the tensor ones (i.e. gravitational waves) remain undetected, the scalar perturbations are observed as temperature fluctuations in the CMB radiation \cite{refId0}. This outstanding prediction makes inflation very appealing as a scenario for the early universe and opens up to many theoretical and observational challenges.

 In this paper we aim at studying the inflationary phase of the universe with a lattice simulation. Many lattice simulations have been developed to study the reheating phase at the end of inflation \cite{Khlebnikov_1996,Prokopec_1997,latticeeasy,Frolov_2008,hlattice,Sainio_2012,Child_2013,Easther_2010,Lozanov_2020,figueroa2021cosmolattice}, where the inflaton decays transferring its energy to the thermal hot bath of the early universe. In the case of reheating lattice simulations are needed because of the non-linear physics involved.

 We will use a lattice simulation to predict the well-known results of some simple single-field models of inflation, particularly for the scalar field perturbation. In these standard models there is no need to use a lattice simulation to study the dynamics of the inflaton, which is well understood with linear perturbation theory. However, this first step is needed to understand the technical aspects involved in simulating the inflationary universe on a discrete lattice. This sets the basis to future work on using lattice simulations to study more complicated inflationary models. Indeed, as we will discuss in \cref{sec:conclusion}, there are many examples of non-standard inflationary models that involve non-linear physics, and where a lattice simulation might be useful.
 
  We will show which aspects need to be taken into account to reproduce the nearly-scale invariant power spectrum of inflaton perturbations for the simplest slow-roll single-field inflationary model. Such technical aspects include, for example, taking into account the modified dispersion relation induced by the finite grid spacing on the modes propagating on the lattice.  Our simulation is written in C++ and it is OpenMP parallelized. It is inspired on LATTICEEASY \cite{latticeeasy}, a lattice code for evolving fields in expanding space-times that has been developed to study the reheating phase of the universe.

The paper is organized as follows. In \cref{sec:inflation} we start with a quick review of inflation to set the notation and recap the equations needed in the rest of the work. In \cref{sec:lattice} we introduce the lattice simulation and we write explicitly the discrete version of the equations of motion. In \cref{sec:fourier} we discuss the difference between continuous and discrete dynamics in Fourier space, and how we need to take it into account to reproduce the results from the linear theory. In section \cref{sec:results} we show the results of the simulation on two simple examples of single-field inflation. Finally, in \cref{sec:conclusion} we summarize the conclusions and implications of this work.

\section{Single field Inflation}
\label{sec:inflation}
We start with a quick recap of inflation in order to present the notation and the equations that will be used in the rest of this work. A more detailed introduction to inflation can be found, for example, in \cite{Baumann:2009ds}. In the standard inflationary picture the action of the early universe is the following:
\begin{equation}
\label{eq:action}
S=\int d^4x \text{ } \sqrt{-g}\biggl(\frac{1}{2}M^2_{\rm Pl}\mathcal{R}-\frac{1}{2}g^{\mu\nu}\partial_\mu\phi\partial_\nu\phi-V(\phi)\biggr),
\end{equation} 
where $\phi$ is the inflaton and $V(\phi)$ its potential. From now on we set $M_{\rm Pl}=1$. In the usual perturbation theory approach, both the scalar field and the metric $g_{\mu\nu}$ are split into a homogeneous background quantity plus a space dependent perturbation. As we will see in \cref{sec:lattice}, this is not true for the lattice simulation, where the inflaton is evolved as a whole without any splitting and the metric is left unperturbed. In this section, however, we still split the inflaton in a background quantity plus perturbations  $\phi(\vec{x},t)=\bar{\phi}(t)+\varphi(\vec{x},t)$ and discuss them separately. This is only done to illustrate the well established results of linear perturbation theory.


Here and in the rest of the paper, the background metric is assumed to be the spatially-flat Friedman-Lemaitre Robinson-Walker metric, with line element:
\begin{equation}
\label{eq:FLRW}
ds^2=g_{\mu\nu}dx^\mu dx^\nu= -dt^2+a^2(t)\left({dr^2}+r^2d\Omega^2\right)=a^2(\tau)\left(-d\tau^2+{dr^2}+r^2d\Omega^2\right),
\end{equation}
where we set $c=1$. 
\subsection{Background evolution}
\label{sec:background}
The expansion of the universe is governed by the Friedman equations:
\begin{equation}
\label{eq:frieds}
H^2=\left(\frac{\dot{a}}{a}\right)^2=\frac{\rho}{3}, \quad\quad \frac{\ddot{a}}{a}=-\frac{1}{6}(\rho+3 p),
\end{equation}
where $\rho$ and $ p$ are the background energy density and pressure of the scalar field, and the dots represent derivatives in cosmic time $t$. At the background level they are written as $\rho = \dot{\bar{\phi}}^2/2+V$ and $ p = \dot{\bar{\phi}}^2/2-V$. From the second Friedman equation we can see that if $\text{ } \dot{\bar{\phi}}^2/2\ll V$ the universe undergoes an accelerated expansion. This is usually called \textit{slow roll} condition. The rate of acceleration is described by the following parameter:
\begin{equation}
	\varepsilon=-\frac{\dot{H}}{H^2}.
\end{equation}
The universe expands in an accelerated way if $\epsilon <1$. Typically, this parameter is much smaller than 1 during slow-roll inflation $\epsilon\ll1$, and approaches 1 at the end of inflation. Varying the action \eqref{eq:action} with respect to the field yields to the Klein-Gordon equation:
\begin{equation}
\label{eq:KG}
\partial_\tau^2\bar\phi+2\mathcal{H}\partial_\tau{\bar\phi}=-a^2\frac{\partial V(\bar\phi)}{\partial{{\bar\phi}}},
\end{equation}
where $\mathcal{H}=a^{-1}\partial_\tau a$. This equation will determine the motion of the background value of the inflaton on the inflationary trajectory.

\subsection{Quantum perturbations}
As mentioned in the introduction, the quantum fluctuations of the inflaton are responsible for the fluctuations of the CMB. In this framework the inflaton is promoted to a quantum operator:

\begin{equation}
\label{eq:quantization}
\varphi(\vec{x},\tau)=\int\frac{d^3\vec{k}}{(2\pi)^{3/2}}\Bigl[{a}_{\vec{k}}{\varphi}(\vec{k},\tau)e^{i\vec{k}\cdot\vec{x}}+{a}_{\vec{k}}^\dagger{\varphi}^\ast(\vec{k},\tau) e^{-i\vec{k}\cdot\vec{x}}\Bigr],
\end{equation}
where $a$ and $a^\dagger$ are the creation and annihilation operators satisfying $[a_{\vec{k}},a^\dagger_{\vec{k}^\prime}]=\delta(\vec{k}-\vec{k}^\prime).$ Varying the action \eqref{eq:action} at second order in perturbation leads to the Mukhanov-Sasaki equation for field fluctuations:
\begin{equation}
\label{eq:MS}
	\partial^2_\tau v +(k^2+m^2_{\rm eff}(\tau))v=0,
\end{equation}
where $v(\vec{k},\tau)=a\varphi(\vec{k},\tau)$ is the Mukhanov variable\footnote{The actual expression for the Mukhanov variable is $v=a(\varphi-\dot{\bar{\phi}} C/H)$, where C is the scalar perturbation of the metric. We work here in the flat gauge $C=0$. The most common choice, however, is working in the co-moving gauge $\varphi=0$. We choose not to work in the co-moving gauge in order to make the connection with the lattice simulation more intuitive.} and $k=|\vec{k}|$. This equation is an harmonic oscillator with a time dependent mass term:
\begin{equation}
	m^2_{\rm eff}(\tau)=-\frac{H}{ \dot{{\bar\phi}}}\partial_\tau^2(\dot{{\bar\phi}}/H).
\end{equation}
The time dependence of the mass introduces a formal ambiguity in choosing the physical vacuum state of the theory. This ambiguity is solved by choosing the so called Bunch-Davies vacuum, which is the ground state of the inflaton in Minkowski space corresponding to the asymptotic past of the theory. This vacuum is associated to the mode functions:
\begin{equation}
\label{eq:BD}
v(\vec{k},\tau)=a{\varphi}(\vec{k},\tau)=\frac{1}{\sqrt{2\omega_k}}e^{- i \omega_{k}\tau},\quad\quad\quad \omega_{k}^2=k^2+m^2,
\end{equation}
where $m$ is the mass of the inflaton. These mode functions serve as the initial conditions for \cref{eq:MS}, which can be solved numerically together with the background equations of \cref{sec:background} to determine the evolution of field fluctuations at linear order. In the case of a free massive scalar field of mass $m$, the effective mass appearing in the Mukhanov-Sasaki equation simplifies to:
\begin{equation}
\label{eq:MSmassive}
m^2_{\rm eff}=m^2a^2-\frac{2}{\tau^2}.
\end{equation}
In this case we can write an analytical solution to the Mukhanov-Sasaki equation with the initial conditions given by \cref{eq:BD} in the asymptotic past $\tau=-\infty$. The solution can be written as:
\begin{equation}
\label{eq:solution}
v(\vec{k},\tau)=a{\varphi}(\vec{k},\tau)=\frac{\sqrt{-\pi \tau}}{2}H_\nu^{(1)}(-k\tau),\quad\quad \nu^2=\frac{9}{4}-\frac{m^2}{H^2},
\end{equation}
where $H_\nu^{(1)}$ is the modified Hankel function of the first kind. We can use this solution to write the power spectrum of inflaton perturbations as $P_{\phi}(k)=|{\varphi}(\vec{k},\tau)|^2$. This gives a theoretical prediction for the dimensionless power spectrum of inflaton perturbation:
\begin{equation}
\label{eq:thprediction}
\mathcal{P}_{\phi}(k)=\frac{k^3}{2\pi^2}P_{\phi}(k)=\frac{H^2}{8\pi}(-k\tau)^3|H_\nu^{(1)}(-k\tau)|^2.
\end{equation}
\section{Lattice simulation}
\label{sec:lattice}
The idea behind a lattice simulation is simple and it consists of simulating the dynamics of continuum fields on a finite cubic lattice. The lattice is defined as a collection of $N^3$ points separated by comoving lattice spacing $dx=L/N$, where $L$ is the comoving physical size of the box. To any given field in continuous space, we associate $N^3$ values to each point of the cubic lattice. For the inflaton $\phi$ this corresponds to:
\begin{equation}
	\phi(\vec{x},\tau) \quad\quad \longrightarrow \quad\quad \phi(\tau)_{i_1,i_2,i_3},\quad i_1,i_2,i_3\in\{1,\dots,N\}.
\end{equation}
To enlighten the notation, from now on we will write $\vec{i}\equiv i_1,i_2,i_3$ for discrete vector indices. Here and for the rest of this work, we use periodic boundary conditions on the lattice.

Contrarily to what is done in perturbation theory, in the simulation we do not split in background and perturbation quantities. Indeed, the inflaton on the lattice is evolved altogether using the Euler-Lagrange equations in real space:
\begin{equation}
\label{eq:disc}
\partial_\tau^2\phi_{\vec{i}}+2\mathcal{H}\partial_\tau{\phi}_{\vec{i}}-L[\phi]_{\vec{i}}+a^2\frac{\partial V}{\partial \phi_{\vec{i}}} =0,
\end{equation}
where $L[\phi]$ is a discretized version of the Laplacian operator. We will mainly consider the following second order stencil for the Laplacian operator:
\begin{equation}
\label{eq:discretelaplacian}
L[\phi]_{\vec{i}}=\frac{1}{dx^2}\sum_{\alpha={+1,-1}}\left(\phi_{i_1+\alpha,i_2,i_3}+\phi_{i_1,i_2+\alpha,i_3}+\phi_{i_1,i_2,i_3+\alpha}-3\phi_{i_1,i_2,i_3}\right).
\end{equation}
In \cref{app:diff} we also consider other stencils for the Laplacian, and show how the choice of the Laplacian affects the dynamics of the simulation.

The space-time is evolved neglecting the perturbations of the metric, which is assumed to be the unperturbed FLRW metric \eqref{eq:FLRW}. Indeed, the scale factor appearing in \cref{eq:disc} is evolved using the second Friedman equation:
\begin{equation}
\label{eq:2fried}
\frac{d^2a}{d\tau^2}=\frac{1}{3}(\rho-3p)a^3.
\end{equation}
In principle, metric perturbations can be included in a full numerical GR computation, as we see for example in \cite{East:2015ggf,Clough:2016ymm}. We choose to neglect them because, as it is well known, their coupling with $\phi$ is slow-roll suppressed\footnote{If we assume to be in the spatially flat gauge $\delta g_{ij}=0$, all the couplings between the inflaton and the metric perturbations $\delta g_{0i}$ and $\delta g_{00}$ are slow-roll suppressed.
	This can be seen expanding the action \eqref{eq:action} to second order in perturbations.}.

The equations of motion of the system \eqref{eq:disc}\eqref{eq:2fried} determine the evolution of the inflationary universe and of the inflaton field on the lattice. We solve them  numerically with a second order staggered-leapfrog integrator inherited from LATTICEEASY \cite{latticeeasy}.
As wee can see, the equations of motion for the evolution of the system are purely classical, even if the inflaton is a quantum field. The quantum behavior of the inflaton will be captured by the initial conditions that, as explained in \cref{sec:initialconditions}, are randomly generated over the lattice. In this picture the uncertainty associated to the quantum nature of the field will be replaced by a statistical uncertainty over the different random realizations of the lattice simulation. This semi-classical approximation is common and it constitutes the working assumption of most of the lattice simulations in the context of inflation.

\section{Discrete dynamics in Fourier space}
\label{sec:fourier}
In this paper we are interested in reproducing the results from the well known linear theory of inflaton perturbations. For this reason, even if the lattice simulation evolves fields in real space, we need to explicitly define how we compute quantities in Fourier space. This is crucial in order to implement the initial conditions, which are given in Fourier space, and to correctly interpret the outputs of the code. As we will see in \cref{sec:modifieddr}, this will be non-trivial due to the modified dispersion relation induced by the discretization of space on the modes propagating on the lattice.





 



\subsection{Discrete Fourier transform and reciprocal lattice}
We start with the definition of lattice modes. For a given discrete field $f_{\vec{i}}$ living on the lattice  we define its Discrete Fourier Transform (DFT) in the following way\footnote{The prefactor $dx^3$ in \cref{eq:DFT} is introduced to take into account the physical discretization of space, and it comes from the $d^3x$ appearing in the integral inside the definition of the continuous Fourier transform:	$$\tilde{f}(\vec{k})={(2\pi)^{-3/2}}\int d^3x f(\vec{x})e^{-i\vec{k}\cdot\vec{x}}.$$}:
\begin{equation}
\label{eq:DFT}
\text{ DFT}[f]_{l_1,l_2,l_3}\equiv \tilde{f}_{\vec{l}}=\frac{dx^3}{N^{3}}\sum_{i_1,i_2,i_3}f_{\vec{i}}\text{ }e^{-i\frac{2\pi}{N}\vec{i}\cdot\vec{l}},\quad\quad\quad l_1,l_2,l_3\in{1,\dots,N}
\end{equation}
In our notation $\tilde{f}$ is the Fourier transform of $f$ for any given field. The Fourier fields live on the reciprocal lattice where we associate to each point the following comoving momentum:
\begin{equation}
\label{eq:modes}
\vec{k}_{\text{lat},l_1,l_2,l_3}=\frac{2\pi}{L}(l_1,l_2,l_3)^\text{T}.
\end{equation}
With this definition, we can write the inverse DFT (iDFT) as follows:
\begin{equation}
\label{eq:iDFT}
f_{\vec{i}}=\frac{1}{dx^3}\sum_{l_1,l_2,l_3}\tilde{f}_{\vec{j}}\text{ }e^{+i\frac{2\pi}{N}\vec{i}\cdot\vec{l}}.
\end{equation}

\subsection{The modified dispersion relation}
\label{sec:modifieddr}
In this section we describe the effect of the discretization on the propagation of modes on the lattice. In continuous space, the Fourier transform (FT) of the Laplacian operator for differential equations is quite simple and reads:
\begin{equation}
\nabla^2\phi(\vec{x})\quad\xrightarrow{\text{FT}}\quad-k^2\phi(\vec{k}).
\end{equation}
As it is well known, this relation gets modified on the lattice \cite{press1986numerical}, where we transform the field with the Discrete Fourier Transform (DFT) (as pointed out for example in \cite{hlattice}). It can be easily derived from  \cref{eq:discretelaplacian} and  \cref{eq:iDFT} that:
\begin{equation}
\normalfont L[\phi]_{i_1,i_2,i_3}\quad \xrightarrow{\text{DFT}}\quad-k_{\text{eff},{l_1,l_2,l_3}}^2\tilde{\phi}_{l_1,l_2,l_3} ,
\end{equation}
where we introduced the effective modes $k_{\text{eff}}$ as:
\begin{equation}
\label{eq:keff}
	k_{\text{eff},{l_1,l_2,l_3}} = \frac{2}{dx}\sqrt{\sin^2\left(\frac{\pi l_1}{N}\right)+\sin^2\left(\frac{\pi l_2}{N}\right)+\sin^2\left(\frac{\pi l_3}{N}\right)}.
\end{equation}
Contrarily to what happen in the continuous case, this relation differs significantly from the value of the $k-$modes of the reciprocal lattice of \cref{eq:modes} $	k_{\text{eff},{l_1,l_2,l_3}} \neq k_{\text{lat},l_1,l_2,l_3}$. Indeed, $k_{\text{lat}}$ and $k_{\text{eff}}$ are only equal in the limit $l_1,l_2,l_3\ll N$. Note that the expression for $k_{\rm eff}$ of \cref{eq:keff} depends on the definition of the lattice Laplacian of \cref{eq:discretelaplacian}. A different choice of the numerical stencil for the Laplacian would lead to a different expression for $k_{\rm eff}$.

This effect is quite general and we can interpret it as a modified dispersion relation induced by the discrete spacing on the modes propagating on the lattice. Indeed, if we look at the equation for a free, massless scalar field on the lattice 
\begin{equation}
\partial^2_\tau\tilde{\phi}_{\vec{l}}=-k_{\text{eff},\vec{l}}^2\text{ }\tilde{\phi}_{\vec{l}},
\end{equation}
we can see that modes will propagate with energy $\omega(k_{\rm lat})=k_{\rm eff}\neq k_{\rm lat}$, which is different from the usual $\omega(k)=k$ of continuous space.

In the case of inflation, we can see the effect of the modified dispersion relation by looking at the lattice version of the Mukhanov-Sasaki equation, which is obtained from \cref{eq:MS} by replacing $k$ with $k_{\rm eff}$:
\begin{equation}
\label{eq:discMS}
\partial^2_\tau v_{\vec{l}} +(k_{\text{eff},\vec{l}}^2+m^2_{\rm eff}(\tau))v_{\vec{l}}=0,
\end{equation}
where $v_{\vec{l}}=a\tilde{\phi}_{\vec{l}}\text{ }$ in analogy to the continuous case. We will use this equation to predict the evolution of perturbations in the simulation. 
In a similar way to \cref{sec:inflation}, we can write an exact solution to this equation in the case of a free massive scalar field:
\begin{equation}
\label{eq:discsolution}
\tilde{\phi}_{\vec{l}}(\tau)=\frac{\sqrt{-\pi \tau}}{2a}H_\nu^{(1)}(-k_{\text{eff},\vec{l}}\text{ }\tau).
\end{equation}
From this solution we can write the lattice dimension-less power spectrum as
\begin{equation}
\label{eq:discthprediction}
\mathcal{P}^{\rm(lat)}_{\phi}(k_{\text{lat}})=\frac{H^2}{8\pi}(-k_{\text{lat}}\tau)^3|H_\nu^{(1)}(-k_{\text{eff}}\tau)|^2,
\end{equation}
which differs from \cref{eq:thprediction} by the presence of $k_{\rm eff}$ instead of $k$ inside the Hankel function. We will discuss in detail the consequences of this in \cref{sec:results}. Notice that the modified dispersion relation will also influence the ground state in Minkowski space, i.e. the Bunch-Davies initial conditions, by changing the expression for the $\omega_k$ appearing in \cref{eq:BD}. We will take this into account when discussing the initial conditions in \cref{sec:initialconditions}.

From the discussion above, and in particular from \cref{eq:discMS} and \cref{eq:discsolution}, it follows that the dynamics of the classical Fourier modes on the lattice is equivalent to the one of the continuous (quantum) modes $\varphi(\vec{k},\tau)$ if we interpret the effective modes $k_{\rm eff}$ as the physical modes actually probed by the lattice simulation. It follows that, when computing the dimensionless power spectrum $\mathcal{P}_{\phi}$ from the simulation, we should multiply the dimension-full power spectrum $|\tilde{\phi}_{\vec{l}}|^2$ by $k^3_{\rm eff}$ instead of $k^3_{\rm lat}$ if we want to match the continuous result\footnote{As we will see, this prescription is useful to compare the power spectrum of the simulation with the continuous theory. However, the reader should keep in mind that only the quantity $\mathcal{P}^{\rm(lat)}_{\phi}(k_{\text{lat}})$ of \cref{eq:discthprediction} is related to the actual variance of the field over the $N^3$ points of the lattice.}. In other words, the non-scaled dimension-full power spectrum of the simulation $|\tilde{\phi}_{\vec{l}}|^2$ will not scale as $\sim k_{\rm lat}^{-3}$, as one would naively expect, but it will scale as $\sim k_{\rm eff}^{-3}$. We will discuss this in more detail when looking at the results of the simulation in \cref{sec:results}.

Moreover, this will have consequences on the effective spacial resolution of the simulation. Instead of probing modes up to\footnote{More details about $k_{\rm lat, max}$ can be found in \cref{app:output}.} $k_{\text{lat},\text{max}}=2\pi\sqrt{3} N_{\text{Nyquist}}/L$, where $N_{\text{Nyquist}}=N/2$, it will probe physical modes up to:
\begin{equation}
\label{eq:effmax}
k_{\text{eff},\text{max}} = \frac{2}{dx}\sqrt{3\sin^2\left(\frac{\pi N_{\text{Nyquist}}}{N}\right)}=\frac{2\sqrt{3}}{dx}=\frac{2}{\pi}k_{\text{lat},\text{max}}.
\end{equation}
This means that the effective range of physical modes evolved by the simulation will be reduced by a factor of $2/\pi\simeq0.64$.

As already mentioned, a different definition of the Laplacian would lead to a different expression for $k_{\rm eff}$ and to a different value of $k_{\rm eff,max}$. In \cref{app:diff} we show the comparison between the $k_{\rm eff}$ associated to different stencils and we discuss the consequences on the dynamics of the simulation.


\subsection{Initial field fluctuations on the lattice}
\label{sec:initialconditions}
We now explain how we generate the initial conditions on the lattice taking into account the modified dispersion relation discussed in the last section. The first step is defining the discrete version of \cref{eq:quantization}:
\begin{align}
\label{eq:discquantization}
\phi_{\vec{i}}=\sum_{\vec{l}}\Bigl[{a}_{\vec{l}}\text{ }u_{\vec{l}}\text{ }e^{i\frac{2\pi}{N}\vec{l}\cdot\vec{i}}+{a}_{\vec{l}}^\dagger\text{ }u_{-\vec{l}}\text{ }e^{-i\frac{2\pi}{N}\vec{l}\cdot\vec{i}}\Bigr]=\frac{1}{dx^3}\sum_{\vec{l}}\tilde{\phi}_{\vec{j}}\text{ }e^{+i\frac{2\pi}{N}\vec{i}\cdot\vec{l}}, 
\end{align}
where in the last equality we show the comparison with our definition of Fourier modes $\tilde{\phi}_{\vec{j}}$. We have also introduced the discrete creation and annihilation operators:
\begin{equation}
[a_{\vec{l}},a^\dagger_{\vec{l}^\prime}]=\delta_L(\vec{l},\vec{l}^\prime)=\frac{1}{L^3}\delta(\vec{l},\vec{l}^\prime).
\end{equation}
Here $u_{\vec{l}}$ are the discrete mode functions, whose exact expression is:
\begin{equation}
\label{eq:discmodes}
	u_{\vec{l}}(\tau)=L^{3/2} \frac{\sqrt{-\pi \tau}}{2a}H_\nu^{(1)}(-k_{\text{eff},\vec{l}}\text{ }\tau).
\end{equation}
At the beginning of the simulation the comoving size of the lattice will be smaller than the horizon $L\lesssim aH$. For this reason, \cref{eq:discmodes} will practically reduce to the Bunch-Davies vacuum of \cref{eq:BD} for most of the modes. 
The extra normalization factor $L^{3/2}$ is introduced to correct for the finite volume of space\footnote{More details about this normalization factor can be found in \cref{app:ic}.}. Note that in LATTICEEASY, and in most of the lattice simulations in the context of reheating, the initial fluctuations are usually generated using $k_{\rm lat}$ instead of $k_{\rm eff}$ in \cref{eq:discmodes}. Once the discrete mode functions $u_{\vec{l}}$ are determined, field fluctuations are generated in Fourier space using \cref{eq:discquantization}, initiating every given mode $\tilde{\phi}_{\vec{l}}\text{ }$ as a Gaussian random number with variance $|u_{\vec{l}}|^2$. We review this procedure in \cref{app:ic}.
 
\section{Results of the simulation}
\label{sec:results}
We now proceed showing the results of the simulation. The main model that we consider is a standard slow-roll inflationary potential for the inflaton $V(\phi)=\frac{1}{2}m^2\phi^2$. We will focus on this model to show the differences between continuous and discrete dynamics described in \cref{sec:fourier}. As a further example, we consider also a similar potential but with a step added on top of it. 
All numerical values in this section are given in Planck units $M_{\rm Pl}=1$. More details on how the outputs are computed in the numerical code can be found in \cref{app:output}, where we also discuss energy conservation in \ref{app:energy}. In \cref{app:ic} we discuss in detail how we generate the initial conditions on the lattice.
\subsection{Standard slow roll potential}
\label{sec:resultsslow}
We first discuss the result for a simple harmonic potential for the inflaton
\begin{equation}
V(\phi)=\frac{1}{2}m^2\phi^2,
\end{equation}
 where $m=0.51\cdot10^{-5}$. This value is chosen to roughly match the COBE normalization of the power spectrum of curvature perturbation $\mathcal{P}_{\zeta}\simeq25\cdot 10^{-10}$.
The initial average value of the inflaton is chosen to be $\bar{\phi}_{\rm in}=14.5$. Its velocity is determined solving numerically the background Klein-Gordon equation \eqref{eq:KG} and it is given by $\bar{\phi}^\prime_{\rm in}=-0.8152m$. With these values the universe is in the middle of the inflationary phase, and there are $N_e\simeq53$ e-folds\footnote{We set as a convention $N_e=0$ at the beginning of the simulation.} left before the end of inflation. The system is evolved until $a=10^3$ ($N_e\simeq6.9$) which means that at the end of the simulation we will still be in the inflationary phase.

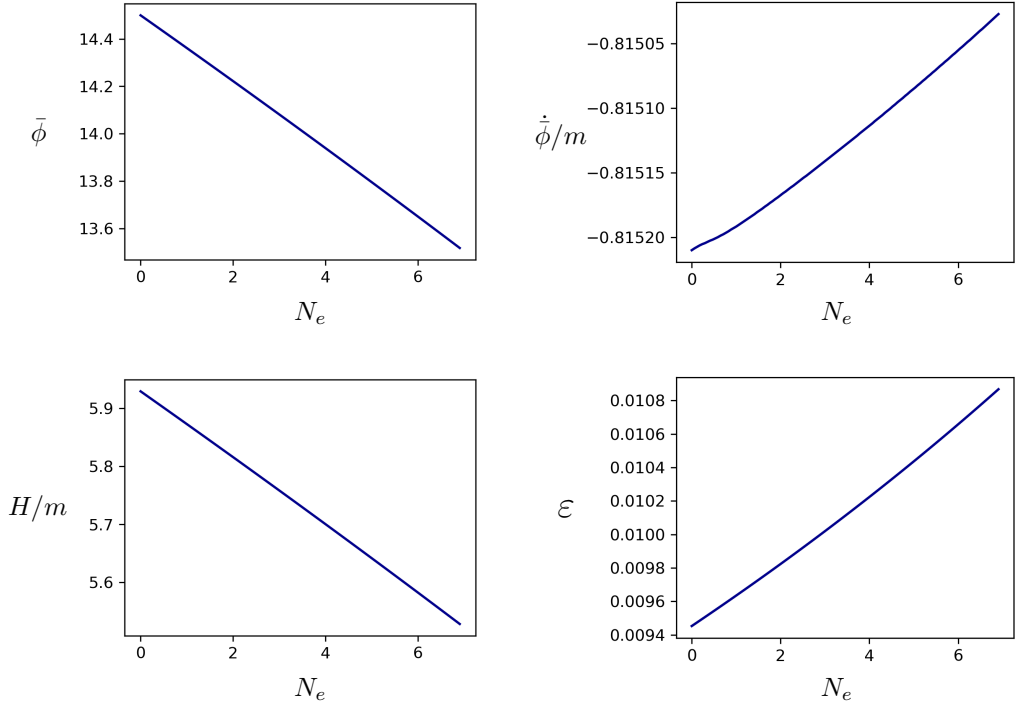
\begin{figure}
	\centering
	
	\begin{tikzpicture}
	\node (img) {\includegraphics[width=6cm]{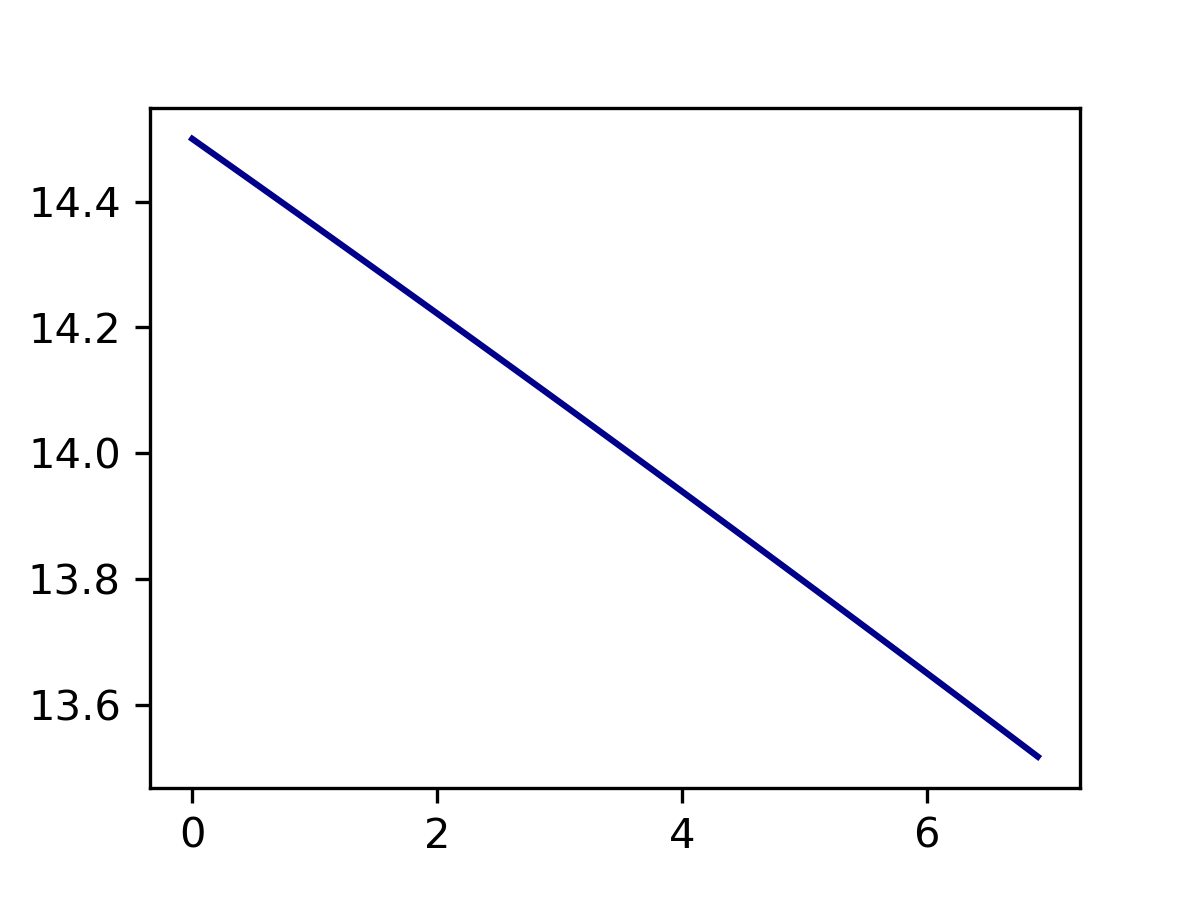}};
	
	\node [rotate=0,text width=0.01cm,align=center] at (-3.5,0){ $\bar{\phi}$};
	\node [text width=0.01cm,align=center] at (0,-2.4){$N_e$};

	\node (img2) at (7,0) {\includegraphics[width=6.4cm]{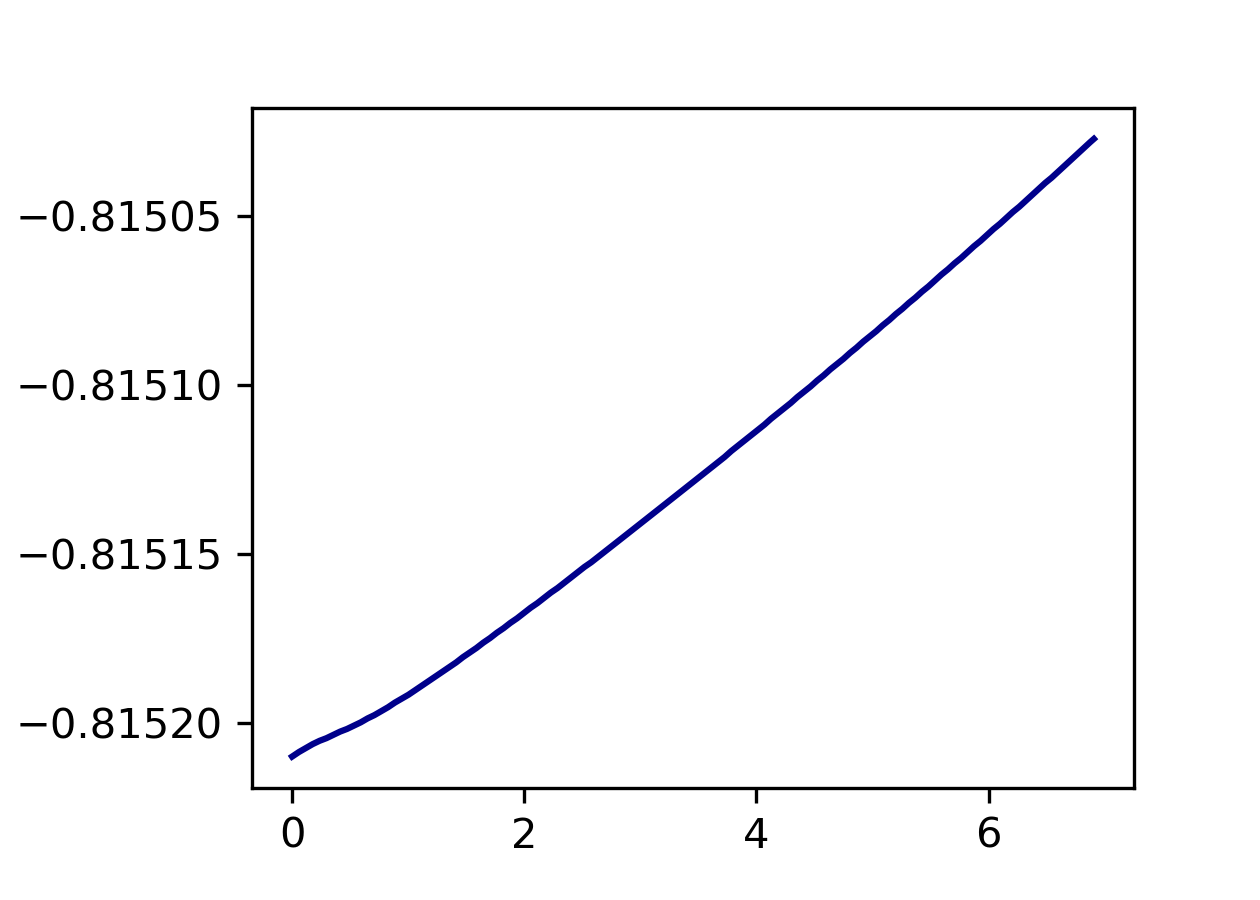}} ;
	
	\node [rotate=0,text width=0.01cm,align=center] at (-3.8+7,0){ ${\dot{\bar{\phi}}}/{m}$};
	\node [text width=0.01cm,align=center] at (0+7,-2.4){$N_e$};

	\node (img3) at (0,-5) {\includegraphics[width=6cm]{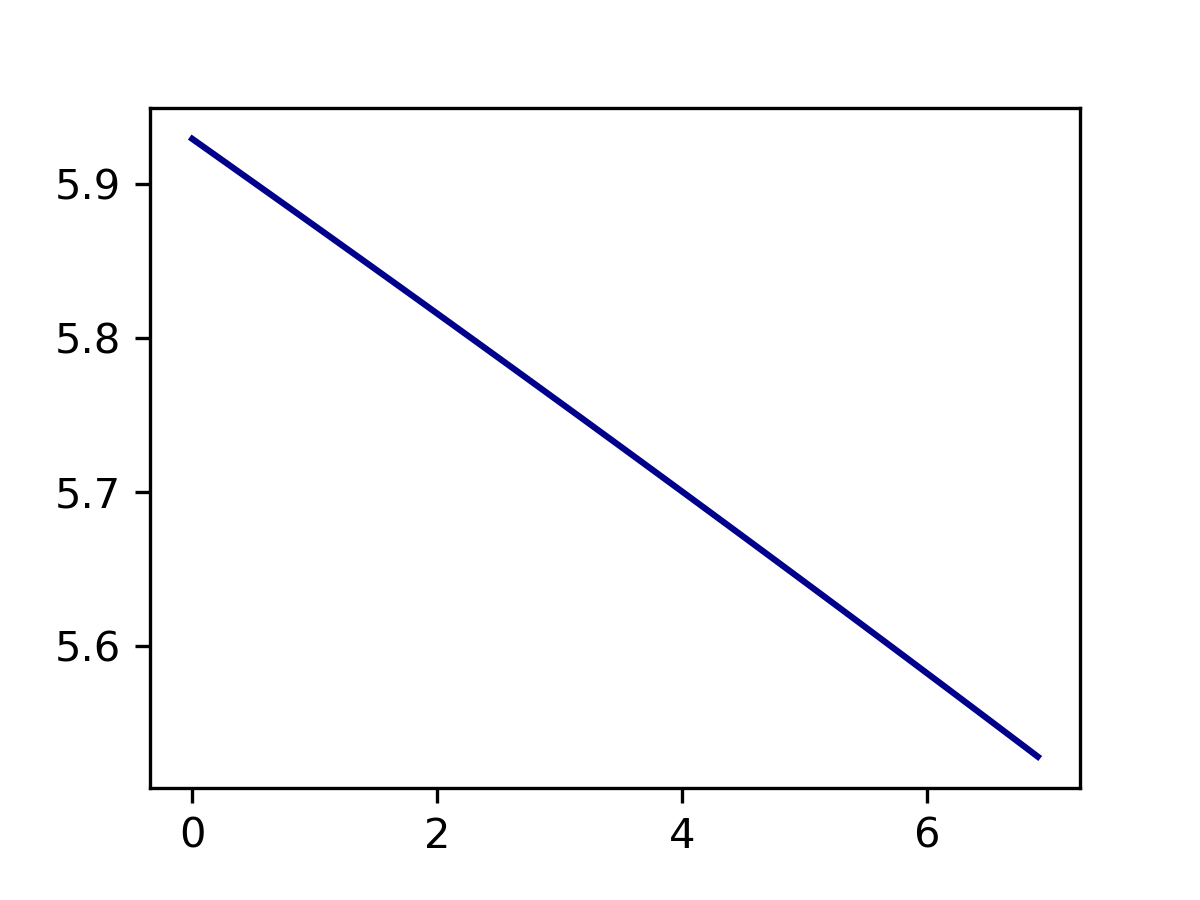}};
	
	\node [rotate=0,text width=0.01cm,align=center] at (-3.8,0-5){ $H/m$};
	\node [text width=0.01cm,align=center] at (0,-2.4-5){$N_e$};

	\node (img4) at (7,-5) {\includegraphics[width=6.4cm]{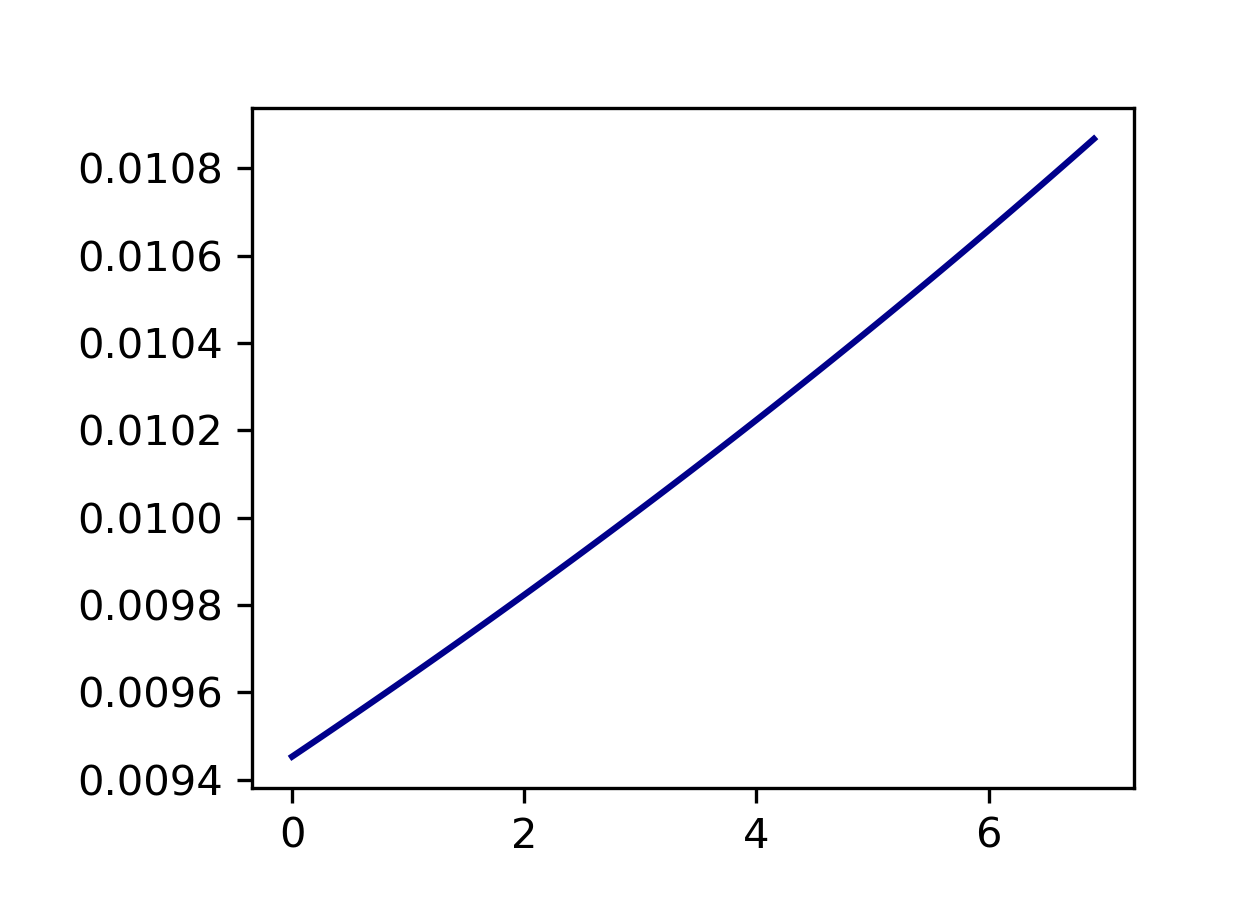}} ;
	
	\node [rotate=0,text width=0.01cm,align=center] at (-3.5+7,0-5){ \Large $\varepsilon$};
	\node [text width=0.01cm,align=center] at (0+7,-2.4-5){$N_e$};

	\end{tikzpicture}

	\caption{Plot of background quantities during the simulation. From top left: the background value of the inflaton $\bar\phi$, its derivative in cosmic time $\dot{\bar{\phi}}$, the Hubble parameter $H$ and the slow-roll parameter $\epsilon$.}
	\label{fig:backgroundvalues}
\end{figure}

In the lattice simulation the inflaton is evolved as a whole, meaning that there is no splitting between background and perturbation quantities. However, we still perform the splitting at the output level in order to compare our results with linear perturbation theory.
 In \cref{fig:backgroundvalues} we show the evolution of the background value of the inflaton $\bar{\phi}$ and its velocity $\dot{\bar{\phi}}$ as functions of the number of e-folds $N_e$. These quantities are computed from the simulation as averages over the $N^3$ points of the lattice. In the same plot, we also show the evolution of $H$ and $\varepsilon$ . The initial bump in the field velocity $\dot{\bar{\phi}}$ is a numerical effect due to the initial stabilization on the inflationary trajectory. From these plots we clearly see that we are in the middle of the inflationary phase, being $\varepsilon \ll1$ and $\dot{{\bar\phi}}\simeq \text{constant} \ll V(\phi)$. 

We now come to the dynamics of field fluctuations and to the importance of the modified dispersion relation discussed in \cref{sec:fourier}. We show results from a run of the code with $L=1.4/m$ and $N^3=128^3$. This translates to:
\begin{equation}
	k_{\rm lat,min}=\frac{2\pi}{L}\simeq 4.49 m \simeq 0.76H_{\rm in}, \quad\quad k_{\rm lat,max}=k_{\rm in} \frac{\sqrt{3}}{2}N\simeq 84.5 H_{\rm in},
\end{equation}
where $H_{\rm in}=\mathcal{H}_{\rm in}$ is the initial value of the Hubble rate. The modes are almost all sub-horizon at the beginning of the simulation. We evolve the system until $a=10^3$, which means that the modes will be all super-horizon at the end of the simulation.

\begin{figure}
	\centering
	
	\begin{tikzpicture}
	\node (img) {\includegraphics[width=7.cm]{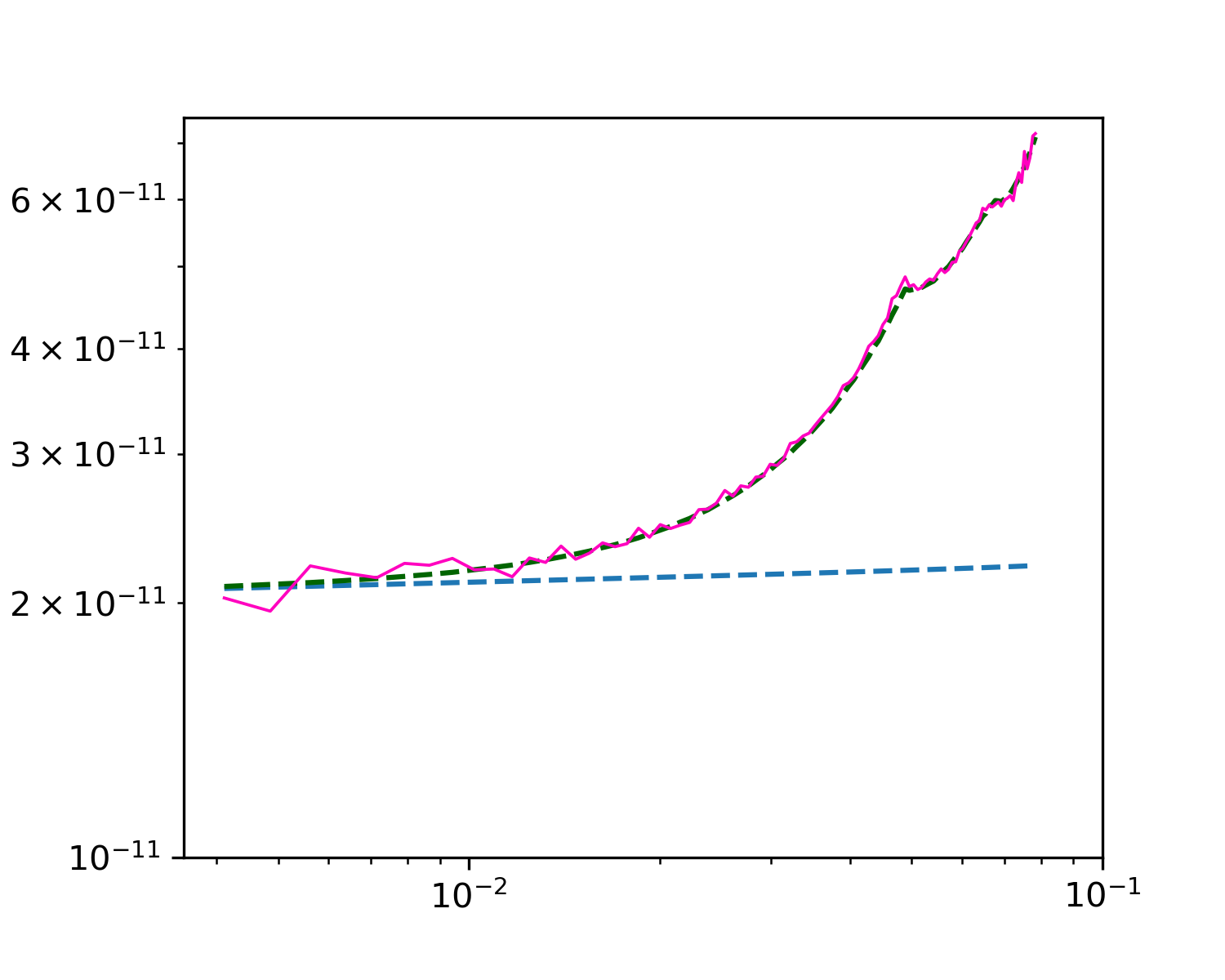}};
	
	\node [rotate=0,text width=0.01cm,align=center] at (-4.6,0){ $\mathcal{P}_{\phi}$};
	\node [text width=0.01cm,align=center] at (0,-3){$k_{\rm lat}/aH$};

	\node (img2) at (7,0) {\includegraphics[width=7.cm]{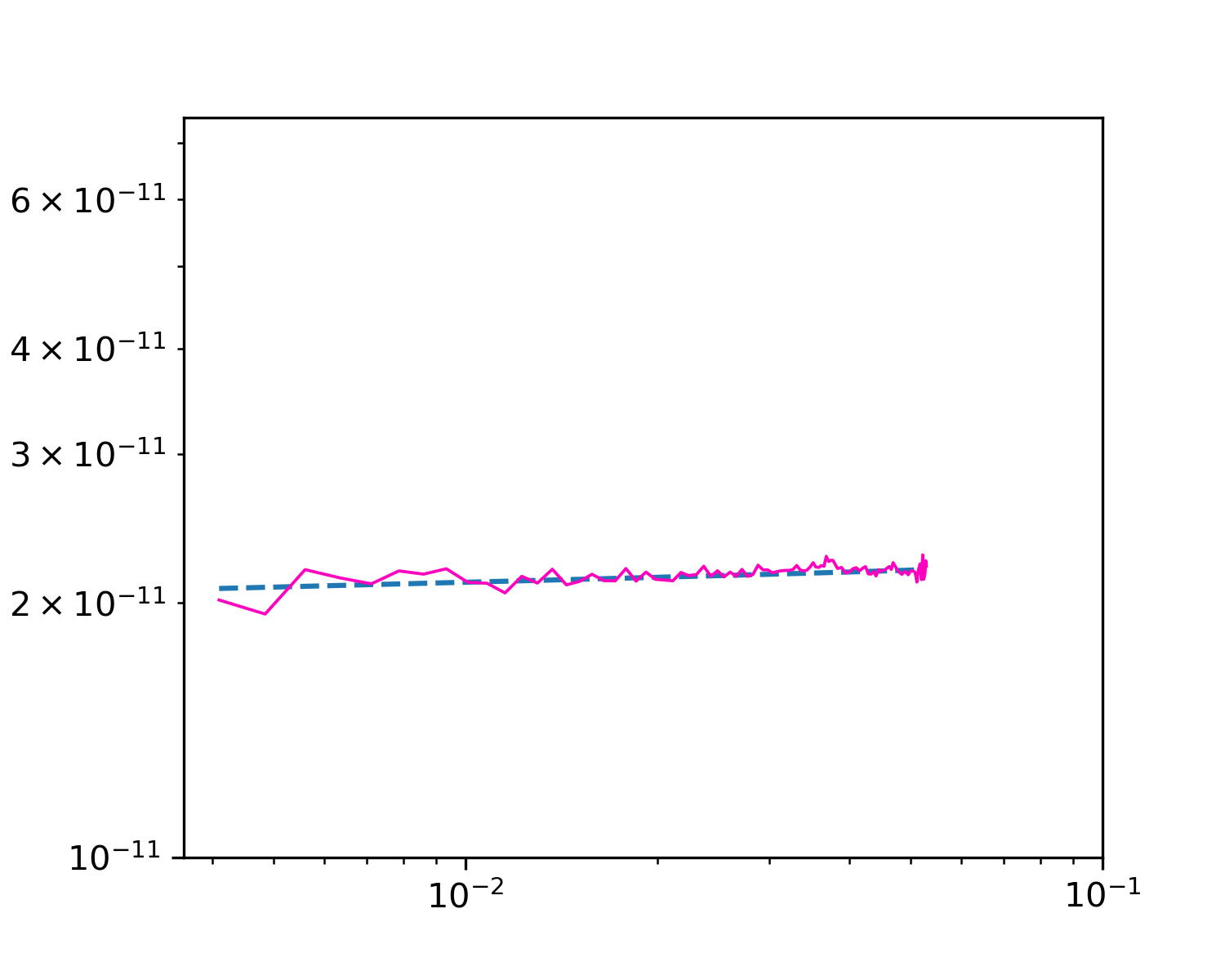}} ;
	
	\node [text width=0.01cm,align=center] at (0+7,-3){$k_{\text{eff}}/aH$};

	\end{tikzpicture}

	\caption{Plot of the final power spectrum computed from the lattice simulation (magenta line) compared to the theoretical prediction of \cref{eq:thprediction} (blue dotted line).On the left panel we show the usual lattice power spectrum, while on the right panel we show the result obtained by taking into account the modified dispersion relation discussed in \cref{sec:modifieddr}.
		The green line in the left panel is the theoretical prediction for the discrete dynamics, as computed from \cref{eq:discthprediction}.}
	\label{fig:finalPS}
\end{figure}
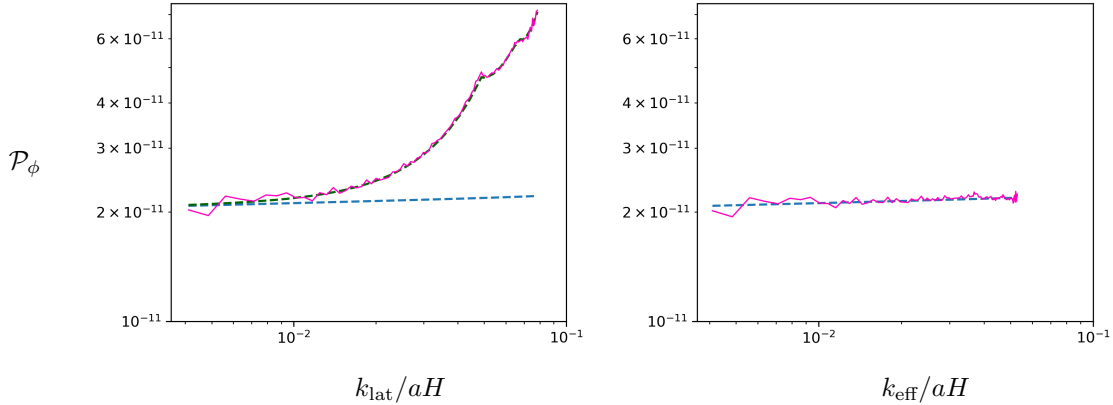

\begin{figure}
	\centering
	
	\begin{tikzpicture}
	\node (img) {\includegraphics[width=7cm]{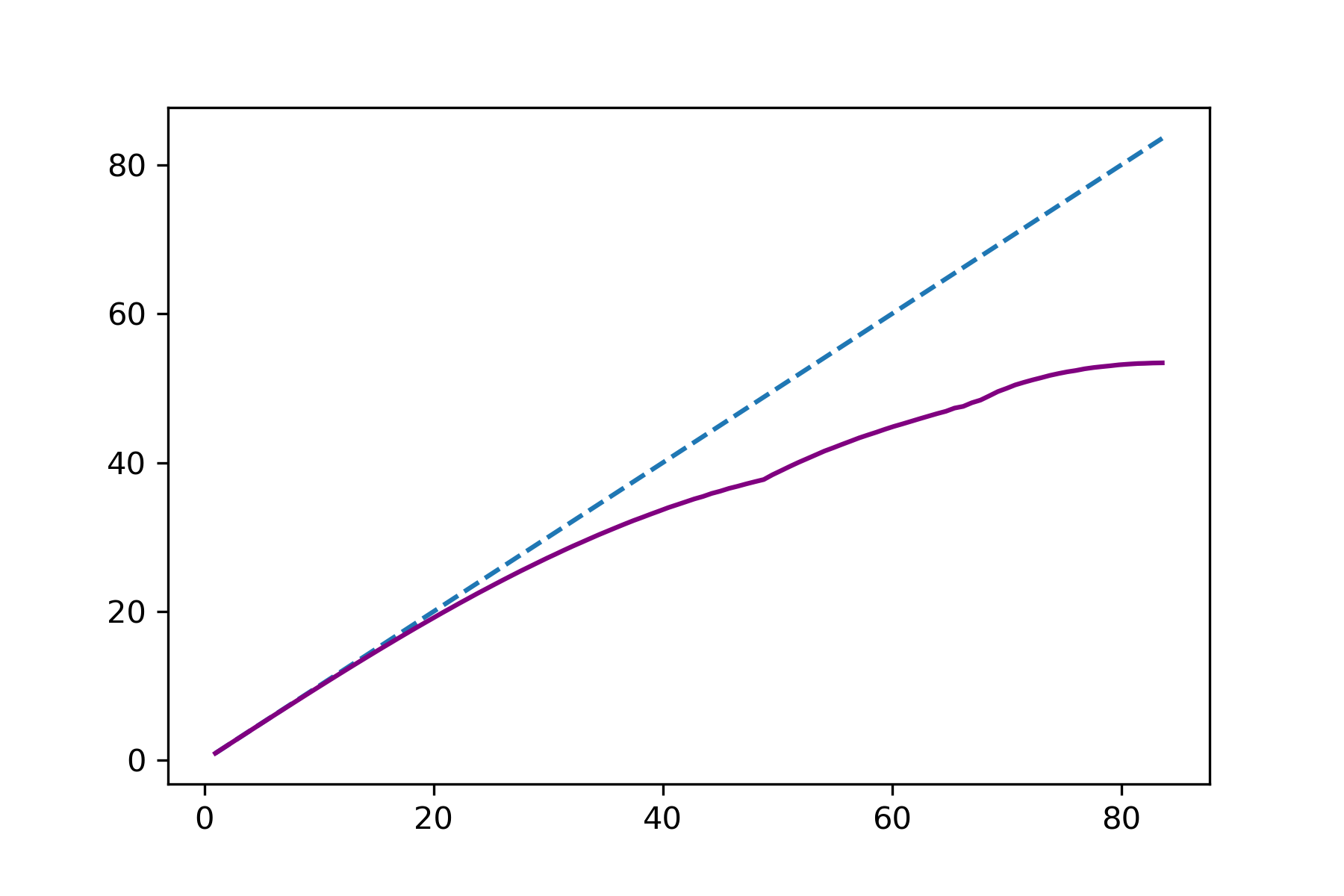}};
	
	\node [rotate=0,text width=0.01cm,align=center] at (-4.3,0){ $k_{\text{eff}}/H_i$};
	\node [text width=0.01cm,align=center] at (-0.3,-2.8){ $k_{\rm lat}/H_i$};

	\end{tikzpicture}

	\caption{The dispersion relation of modes on the lattice. On the y-axis we show $k_{\text{eff}}$ obtained from \cref{eq:keff}, while on the x-axis we show the lattice modes of \cref{eq:modes}. The departure of the diagonal is a manifestation of the modified dispersion relation induced by lattice spacing, as discussed in \cref{sec:modifieddr}.}
	\label{fig:dispersion}
\end{figure}

In the left panel of \cref{fig:finalPS} we show the dimensionless power spectrum of the inflaton $\mathcal{P}_\phi$ at the end of the simulation, plotted against lattice modes $k_{\rm lat}$ of \cref{eq:modes}. We compare the power spectrum computed from the simulation with the theoretical prediction for discrete dynamics of \cref{eq:discthprediction}, which is shown as a green line. From this plot we can see that the discrete power spectrum is quite different from the almost scale-invariant power spectrum of the continuous theory, given by \cref{eq:thprediction} and depicted as a blue line in the plot. This is a manifestation of the different dynamics in Fourier space between the discrete and the continuous case and it is a manifestation of the modified dispersion relation. However, as we have discussed in \cref{sec:modifieddr}, continuous and discrete dynamics are equivalent if we interpret $k_{\rm eff}$ of \cref{eq:keff} instead of $k_{\rm lat}$ as the physical modes probed by the lattice simulation. In \cref{fig:dispersion} we can see a comparison between $k_{\text{eff}}$ and lattice modes $k_{\rm lat}$. The departure from the diagonal is a manifestation of the modified dispersion relation discussed in \cref{sec:modifieddr}. The one dimensional quantities $k_{\rm eff}$ and $k_{\rm lat}$ are obtained averaging \cref{eq:keff,eq:modes} over spherical bins on the lattice\footnote{See \cref{app:output} for more details about the computation of $k_{\rm lat}$ and $k_{\rm eff}$.}.


In the right panel of \cref{fig:finalPS} we show the power spectrum from the simulation computed interpreting $k_{\rm eff}$ as physical modes and we compare it to the theoretical prediction of \cref{eq:thprediction}.
From these plots we see that only in this case we are able to reproduce with precision the nearly scale-invariant spectrum of single-field inflation. 
Note that interpreting $k_{\rm eff}$ as the physical modes will also reduce the resolution in Fourier space, that is computed from \cref{eq:effmax} and it is given by $k_{\text{eff,max}}\simeq 53.50H_{\rm in}$. 
In \cref{app:diff} we also consider results of simulations with different stencils for the discrete Laplacian and compare the corresponding effective momenta.

\begin{figure}
	\centering
	
	\begin{tikzpicture}
	\node (img) {\includegraphics[width=14cm]{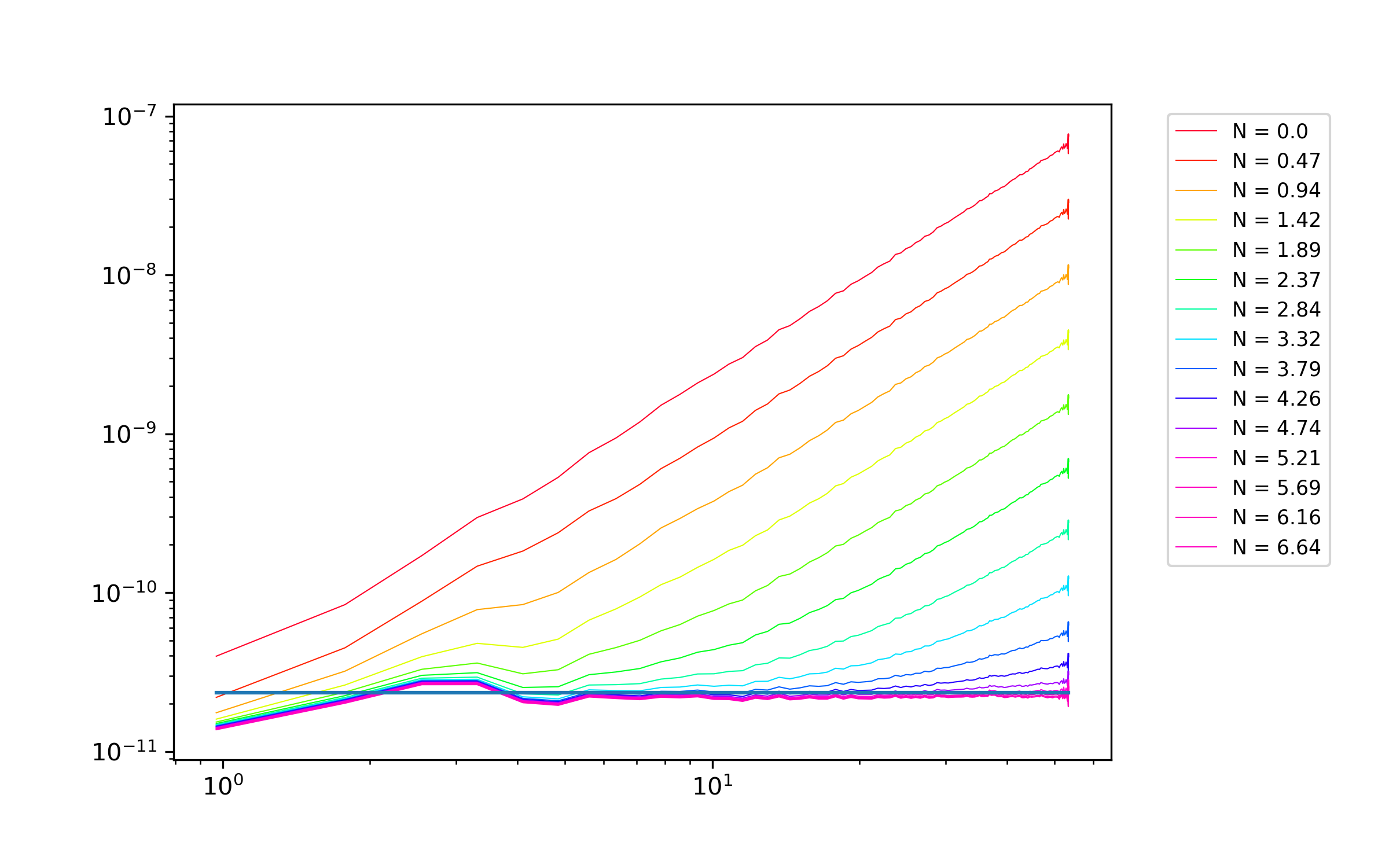}};
	\node [rotate=0,text width=0.01cm,align=center] at (-7,0){ $\mathcal{P}_{\phi}$};
	\node [text width=0.01cm,align=center] at (-.5,-4.2){$k_{\text{eff}}/H_i$};
	
	\end{tikzpicture}

	\caption{The evolution of the power spectrum of inflaton perturbation during the simulation. The colors go from early times (red) to late time (purple). The blue line represents the theoretical prediction for the final power spectrum as computed from \cref{eq:thprediction}.}
	\label{fig:ps}
	
\end{figure}

In \cref{fig:ps} we show the evolution of the power spectrum during the simulation, plotted at different times as a function of physical modes and going from the early-time Bunch-Davies state to the final scale-invariant state.

\subsection{Potential with a step}
\label{sec:resultsstep}
As a further example, in this section we show the results of the code for a model with potential:
\begin{equation}
	V(\phi)=\frac{1}{2}m^2\phi^2\left[1+ s\tanh\left(\frac{\phi-\phi_{\rm step}}{d}\right) \right].
\end{equation}
This potential is analogous to the harmonic potential of the last section but with a step localized at $\phi_{\rm step}$. This model has been studied in \cite{Adams_2001}, where the authors show that the presence of the step causes oscillations in the power spectrum of scalar perturbations. Here we show results for the same parameters of the last section. The only difference here is that we use $L=0.6/m$ as comoving size of the box, that corresponds to $k_{\text{eff,max}}\simeq124.84 H_{\rm in}$. Moreover, we have three extra parameters $s$, $d$ and $\phi_{\rm step}$. We choose $\phi_{\rm step} = 14.35$, and we run the simulation with different values of $s$ and $d$.
\begin{figure}
	\centering
	
	\begin{tikzpicture}
	\node (img) {\includegraphics[width=6cm]{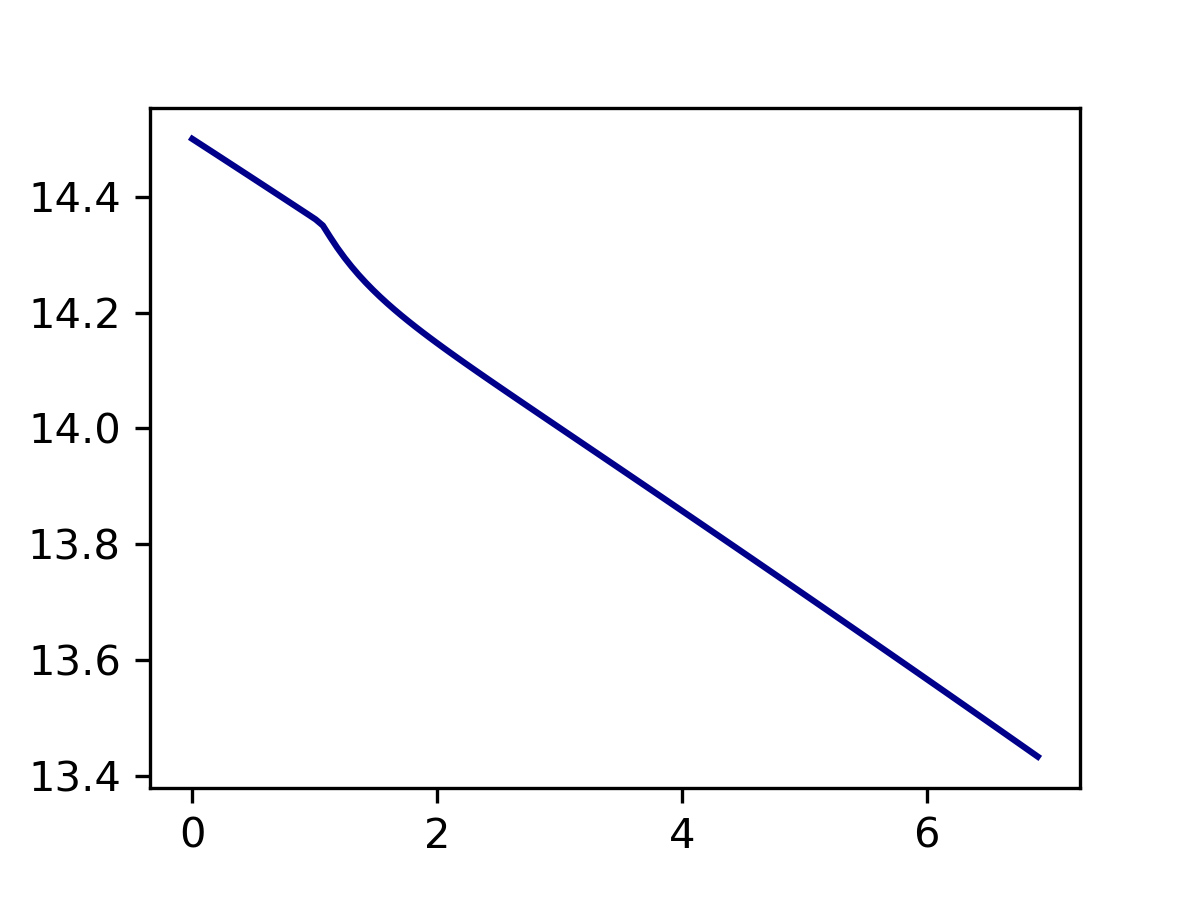}};
	
	\node [rotate=0,text width=0.01cm,align=center] at (-3.5,0){ $\bar{\phi}$};
	\node [text width=0.01cm,align=center] at (0,-2.4){$N_e$};

	\node (img2) at (7,0) {\includegraphics[width=6.4cm]{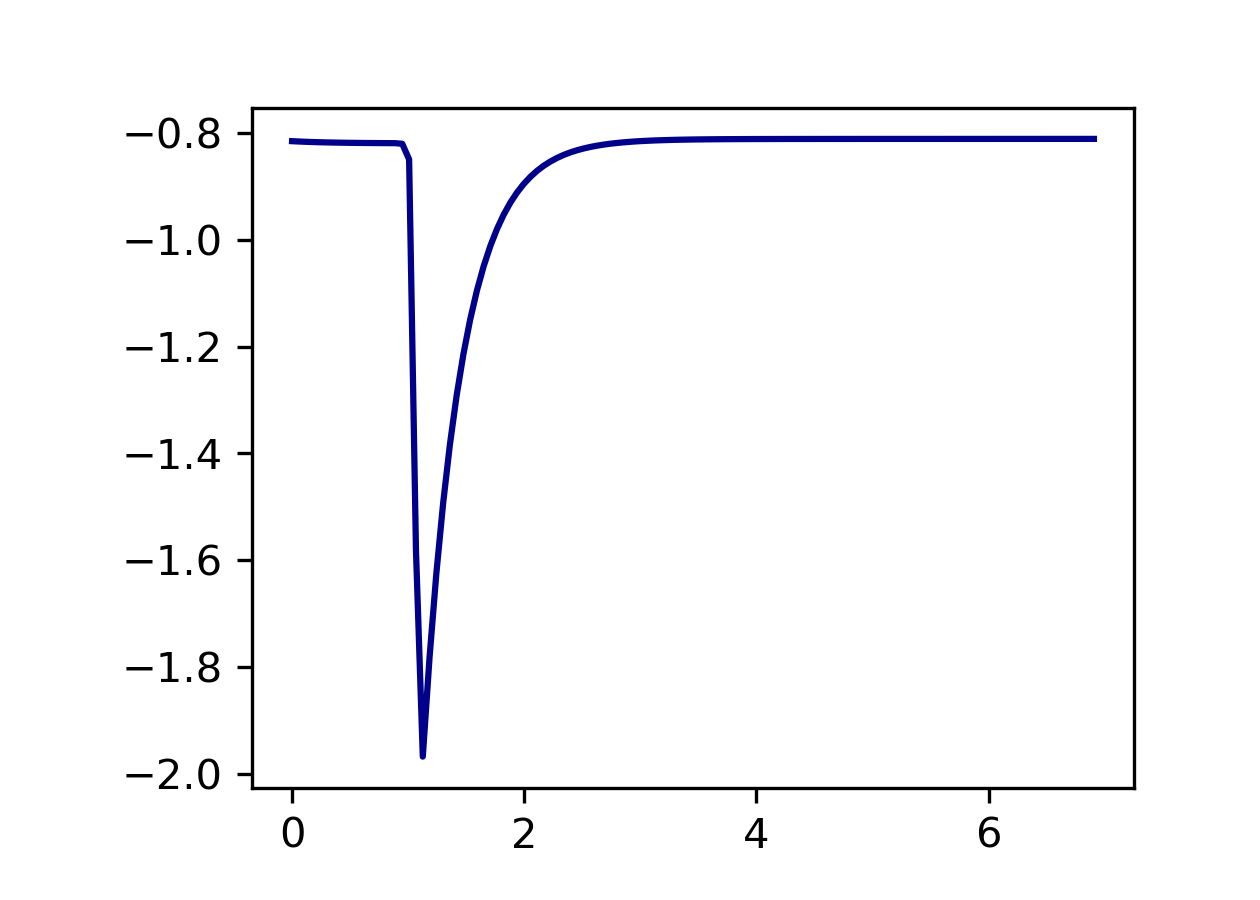}} ;
	
	\node [rotate=0,text width=0.01cm,align=center] at (-3.6+7,0){ ${\dot{\bar{\phi}}}/{m}$};
	\node [text width=0.01cm,align=center] at (0+7,-2.4){$N_e$};

	\node (img3) at (0,-5) {\includegraphics[width=6cm]{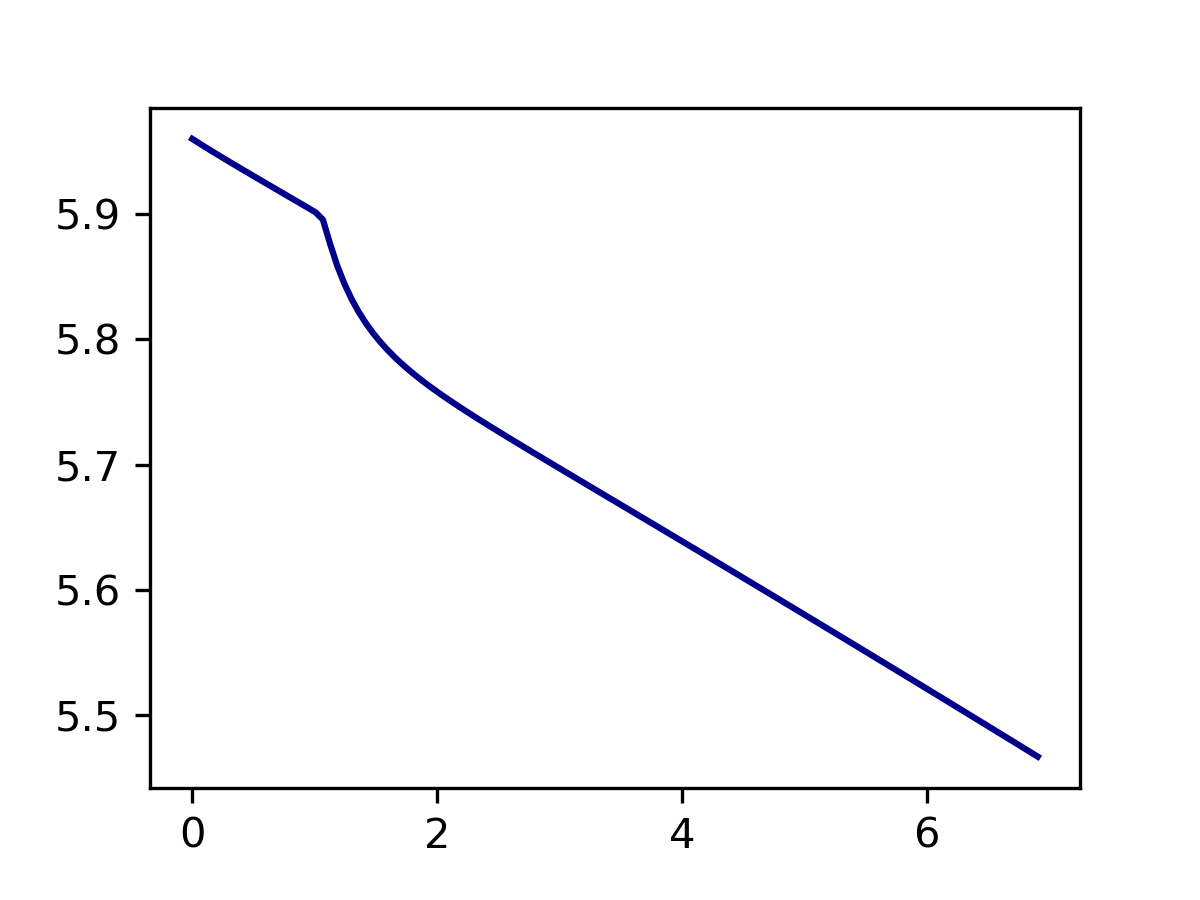}};
	
	\node [rotate=0,text width=0.01cm,align=center] at (-3.8,0-5){ $H/m$};
	\node [text width=0.01cm,align=center] at (0,-2.4-5){$N_e$};

	\node (img4) at (7,-5) {\includegraphics[width=6.4cm]{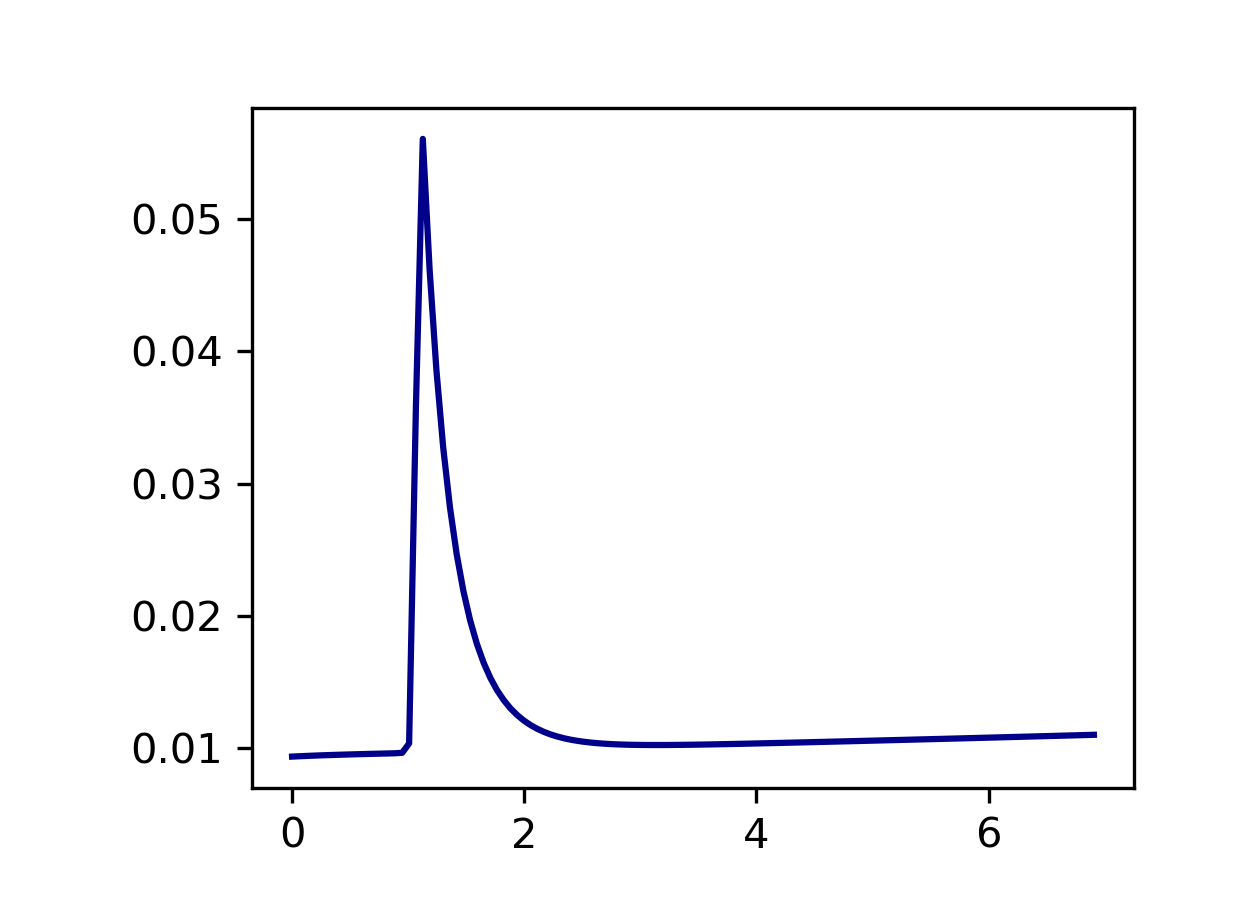}} ;
	
	\node [rotate=0,text width=0.01cm,align=center] at (-3.5+7,0-5){ \Large $\varepsilon$};
	\node [text width=0.01cm,align=center] at (0+7,-2.4-5){$N_e$};

	\end{tikzpicture}

	\caption{Plot of background quantities during the simulation for a potential with a step with $s=0.01$, $d=0.005$ and $\phi_{\rm step}=14.35$. From top left: the background value of the inflaton $\bar\phi$, its derivative in cosmic time $\dot{\bar{\phi}}$, the Hubble parameter $H$ and the slow-roll parameter $\epsilon$.}
	\label{fig:backgroundvalues_step}
\end{figure}
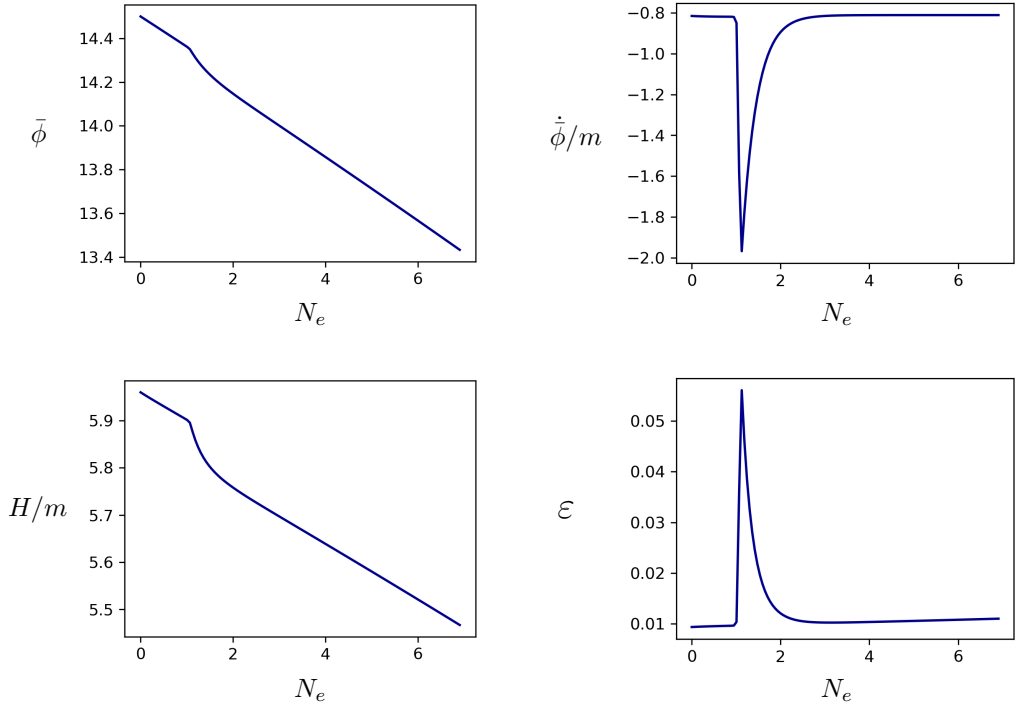

In \cref{fig:backgroundvalues_step} we show the background quantities in the case $s=0.01$, $d=0.005$. We can see here that the step of the potential causes a bump in all the background quantities, but without changing significantly the slow-roll dynamics of the inflaton. Indeed, $\varepsilon$ is still smaller than 1 during the simulation and the departure of the inflaton from the slow-roll trajectory is small. 

\begin{figure}
	\centering
	
	\begin{tikzpicture}
	\node (img) {\includegraphics[width=12cm]{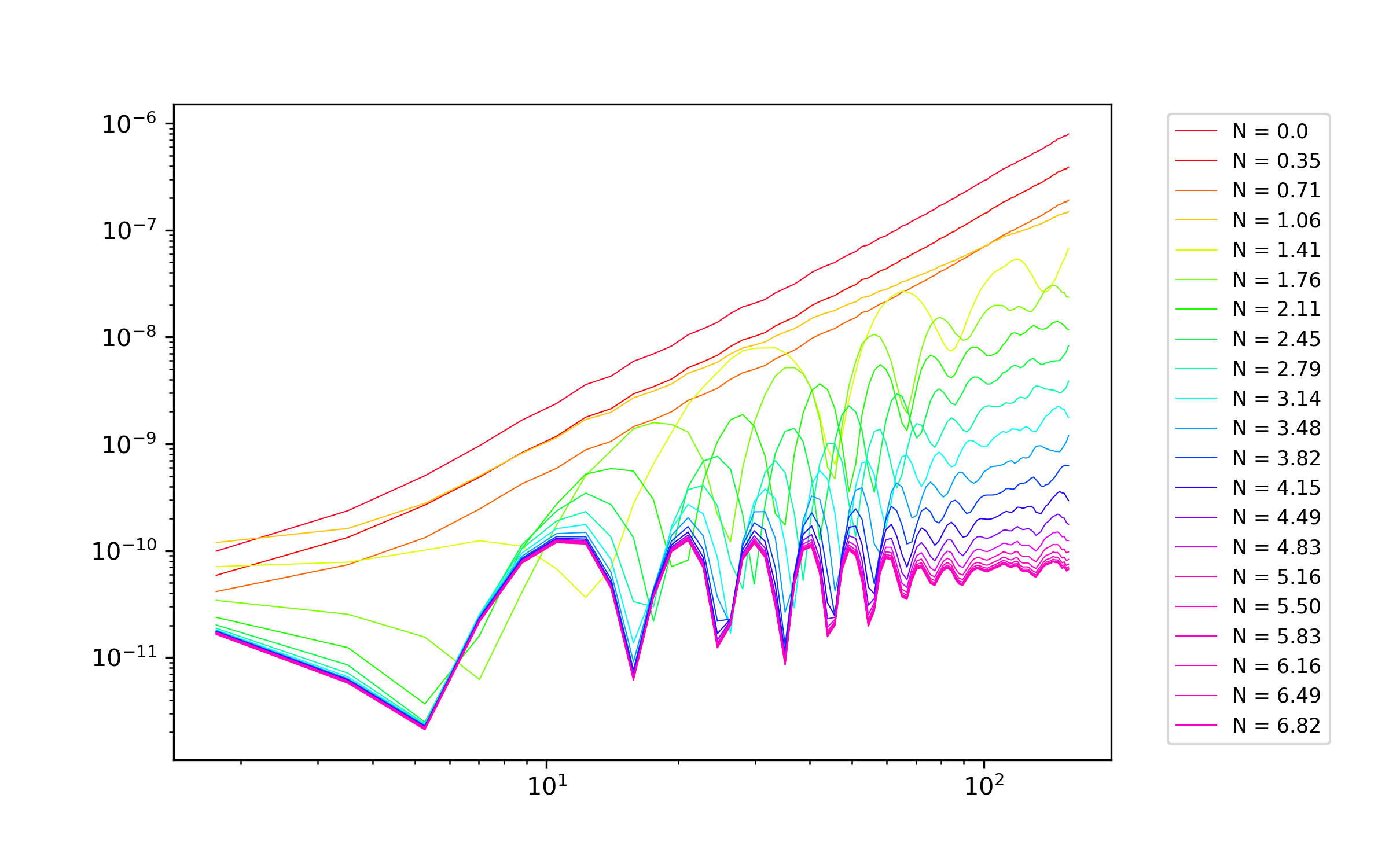}};
	\node [rotate=0,text width=0.01cm,align=center] at (-6.3,0){ $\mathcal{P}_{\phi}$};
	\node [text width=0.01cm,align=center] at (-.5,-3.5){$k_{\text{eff}}/H_i$};

	\end{tikzpicture}

	\caption{The evolution of the power spectrum of inflaton perturbation during the simulation for the step potential. The colors go from early times (red) to late time (purple). We show the result for $s=0.01$ and $d=0.005$.}
	\label{fig:ps_step}
	
\end{figure}

In \cref{fig:ps_step} we show the evolution of the power spectrum during the simulation for $s=0.01$ and $d=0.005$. Here we can clearly see that the presence of the step introduces oscillations in the power spectrum.

\begin{figure}
	\centering
	
	\begin{tikzpicture}
	\node (img) {\includegraphics[width=7.cm]{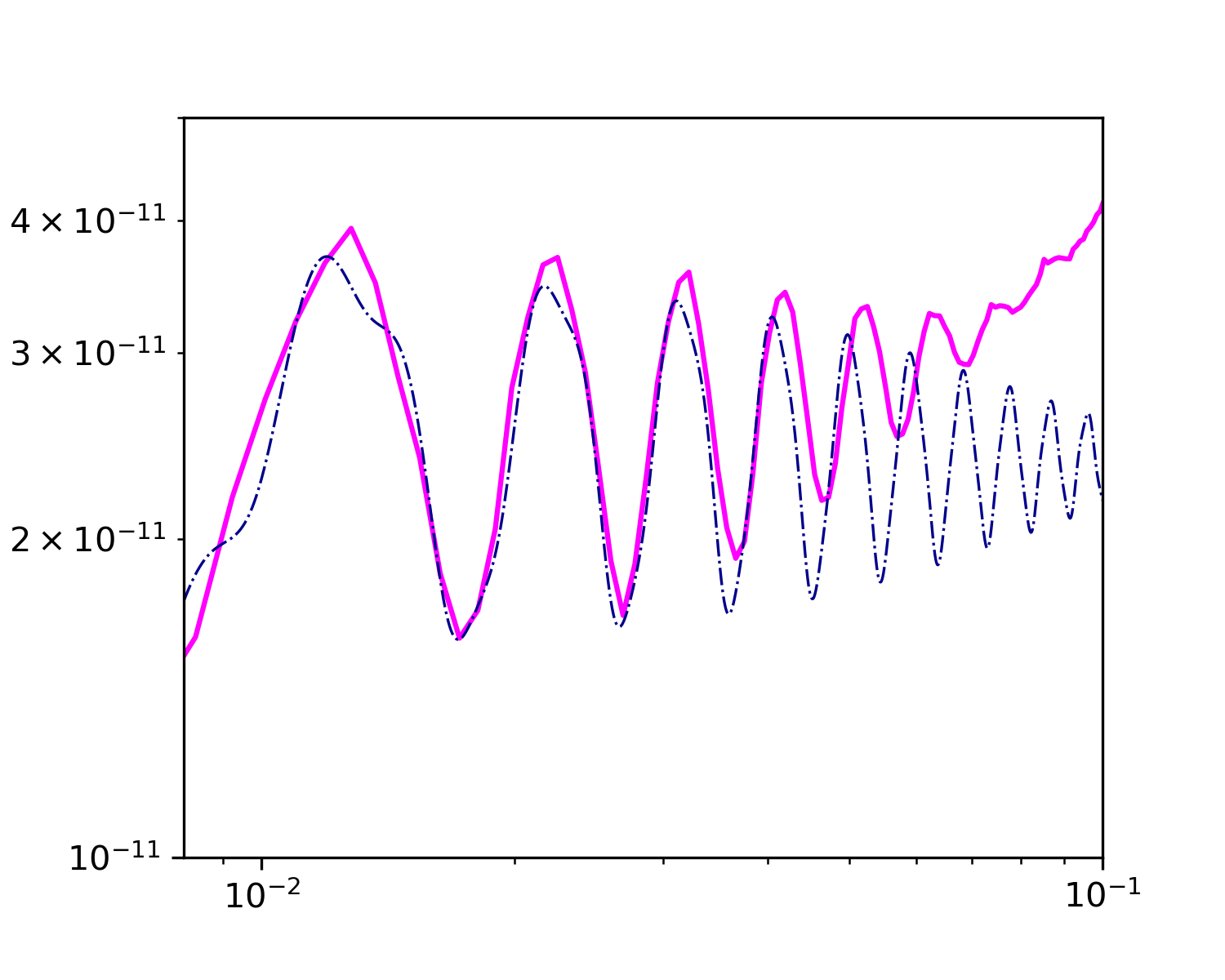}};
	
	\node [rotate=0,text width=0.01cm,align=center] at (-4.5,0){ $\mathcal{P}_{\phi}$};
	\node [text width=0.01cm,align=center] at (0,-2.7){$k_{\rm lat}/aH$};

	\node (img2) at (7,0) {\includegraphics[width=7.cm]{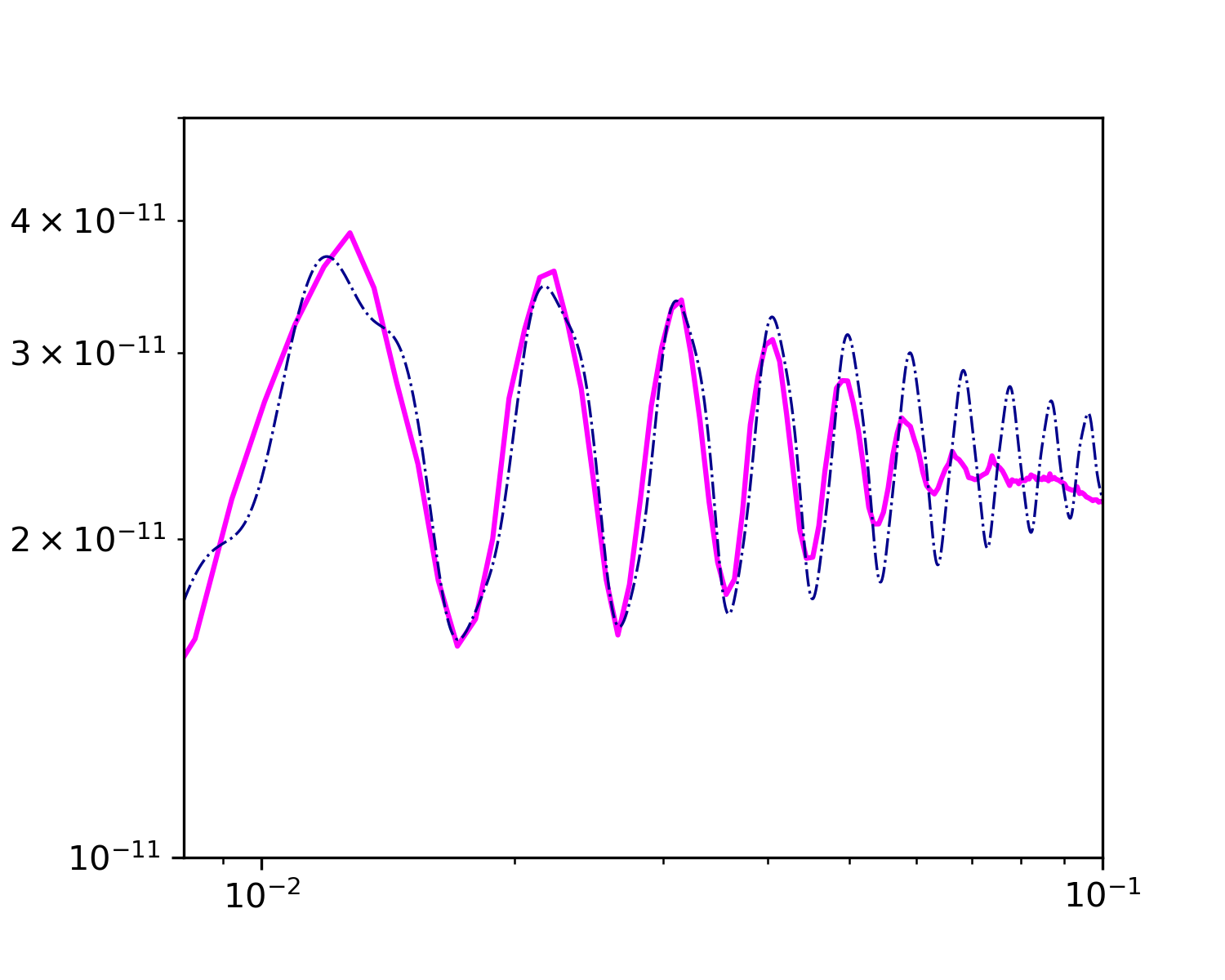}} ;
	
 	\node [text width=0.01cm,align=center] at (0+7,-2.7){$k_{\text{eff}}/aH$};

	\end{tikzpicture}

	\caption{Plot of the final power spectrum computed from the lattice simulation (magenta line) compared with the theoretical prediction computed with a linear code that solves numerically the Mukhanov-Sasaki \cref{eq:MS} (blue dotted line). The step parameters are $s=0.001$ and $d=0.005$ for this plot. On the right panel we show the result obtained taking into account the modified dispersion relation discussed in \cref{sec:modifieddr}. On the left we show the usual lattice power spectrum.}
	\label{fig:finalPS_step}
\end{figure}
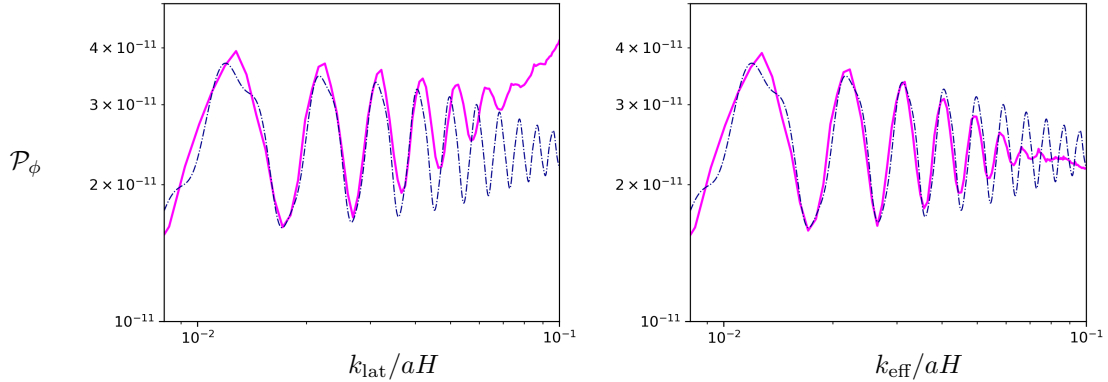

In \cref{fig:finalPS_step} we show the final power spectrum of a simulation run with $s=0.001$, $d=0.005$ and we compare it with the result obtained solving the Mukhanov-Sasaki equation \eqref{eq:MS} with a numerical integrator, which serves as a theoretical prediction.  For this simulation we increased the number of lattice points to $N^3=256^3$ and the box size to $L=1.2/m$ in order to improve the spatial resolution. In the right panel of this figure we show the result obtained by interpreting $k_{\rm eff}$ the physical modes, while on the left panel we show the lattice power spectrum. From the right plot we can see that the matching between the theoretical prediction and the lattice simulation is not perfect, in particular for the largest modes of the simulation. However, the lattice code is able to correctly reproduce the oscillations, that have the same frequency and a similar amplitude compared to the theoretical prediction. 
In this example we can again see that interpreting $k_{\rm eff}$ as the physical modes is important in order to get a sensible result.

	
	




\section{Conclusions and outlook}

\label{sec:conclusion}
In this work we studied some well known models of single-field inflation with a lattice simulation. After introducing the numerical code,
we discussed the effects induced by the discrete lattice that need to be understood in order to reproduce the well known results of single-field inflation for the scalar field perturbation. We have seen that an important role in reproducing the results of the continuous theory is played by the identification of the physical modes that are probed by the lattice simulation. In order to do so, we studied how the finite grid spacing influences the dispersion relation of modes propagating on the lattice.

Taking into account the modified dispersion relation was necessary in order to correctly interpret the power spectra computed from the simulation and to properly initialize the scalar field on the lattice. This allowed us to match the results from the linear theory in two cases. The first was a simple harmonic potential for the inflaton, while the second was a similar potential but with a step added on top of it. 

As we already stated, a non-linear lattice simulation is not needed to understand the physics of these simple models of inflationary potential, that lies well within the regime of validity of linear perturbation theory. However, there are many non-standard models of inflation that involve non-linear processes and that have been studied in the literature because of their interesting phenomenology. Examples are the geometrical destabilization of inflation \cite{Renaux_Petel_2016}, or inflationary models with abelian \cite{Anber_2006,Anber_2010,Barnaby_2011_Large, Barnaby_2011, Anber_2012} and non-abelian \cite{Maleknejad_2011,maleknejad2013gaugeflation,Adshead_2012,Adshead_2013,maleknejad2021su2r} gauge fields. These models potentially manifest strong backreaction effects, that invalidate the use of perturbation theory \cite{Renaux_Petel_2016,Ferreira:2015omg,Peloso_2016,Maleknejad_2019,Lozanov_2019}. For this reason, lattice simulations might turn out to be an important tool in order to compute observables from these kind of models, such as the 2- and 3-point functions of scalar (and tensor) perturbations. This work constitutes a first step in this direction. Indeed, even if some examples of lattice study during inflation already exist in the literature (see \cite{Barnaby_2009} for example), this is the first time in which a lattice simulation is used to recover the 2-point function of the inflaton in single-field models with such precision. This is a necessary result if one wants to use a lattice simulation to compute observables from more complicated models. Moreover, a precise computation of the 2-point function is a necessary step in developing a code that is able to compute 3-point statistics during inflation, which might be important in many applications.


\section*{Acknowledgements}
This work is supported in part by the Excellence Cluster ORIGINS which is funded by the Deutsche Forschungsgemeinschaft (DFG, German Research Foundation) under Germany's Excellence Strategy - EXC-2094 - 390783311 (AC, EK, JW), and JSPS KAKENHI Grant Number JP20H05859 (EK). The Kavli IPMU is supported by World Premier International Research Center Initiative (WPI), MEXT, Japan. The work of KL is supported in part by the US Department of Energy through grant DESC0015655. We thank anonymous referees for giving us many useful suggestions, which have improved the precision and robustness of the results in this paper significantly.

\appendix

\section{Initial conditions on the lattice}
\label{app:ic}
\subsection{Background quantities}
The initial background values of the inflaton $\bar{\phi}_{\rm in}$ and its velocity $\bar{\phi}^\prime_{\rm in}$ are set by requiring that the universe is in the middle of the inflationary phase. This will depend on the inflaton potential, that we will write explicitly in \cref{sec:results}. The scale factor $a$ is simply set to $1$ at the beginning of the simulation, while its derivative in program time $a^\prime$ is computed from the first Friedman equation \eqref{eq:frieds} using only the background energy density and pressure of the field and neglecting gradient contributions:
$$3a^\prime_{\rm in}=\bar\rho = \frac{	({\bar\phi}^\prime)^2}{2} + V(\bar\phi).$$
 Then, after the field fluctuations are generated, the value of $a_{\rm in}^\prime$ is updated to include the gradient term, computed as a lattice average of $(\partial_i\phi)^2/2$. Note that quantum vacuum sub-horizon fluctuations should not contribute to the Friedmann equations. However, we still include them in generating the initial value of $H$, and this is a consequence of our semi-classical approximation. This important in order to evolve the discrete system in a consistent way. Indeed, neglecting gradient contributions will result in an effective residual curvature in the second Friedmann equation, that we use to evolve the scale factor during the simulation. Moreover, including the gradient term in a consistent way allows us to check that energy is conserved in the discrete system, i.e. to ensure that there are no numerical errors propagating on the lattice during the simulation. More about the energy conservation check can be found in \cref{app:energy}.
 
 
\subsection{Initial fluctuations of the field}
\label{app:icf}
The field fluctuations are generated starting from the expression of the mode functions of \cref{eq:discmodes}. The extra factor of $L^{3/2}$ is a common normalization \cite{latticeeasy,Frolov_2008} introduced to correct for the finite volume of space. In order to understand this, take the two point function of the field:
\begin{equation}
\langle\phi_{\vec{i}}^2\rangle=	\langle 0|\phi_{\vec{i}}^2|0\rangle=\sum_{\vec{l},\vec{l}^\prime}\delta_L(\vec{l},\vec{l}^\prime)u_{\vec{l}}\text{ }u_{-\vec{l}^\prime}=\frac{1}{L^3}\sum_{\vec{l}}|u_{\vec{l}}|^2.
\end{equation}
We can clearly see that this scales as $L^{-3}$ due to the presence of the finite-volume delta function $\delta_L$. If we want the quantity $\langle \phi^2 \rangle$, and the two-point functions in general, to be independent of the physical size of the lattice we have to normalize the mode functions by a factor of $L^{3/2}$.

In our classical simulation we take a statistical point of view, interpreting the quantum creation and annihilation operator as stochastic variables that take different values at each realization. In this picture the creation and annihilation operators of \cref{eq:discquantization} are initiated as:
\begin{equation}
a_{\vec{l}}=e^{i 2\pi Y_{\vec{l}}}\sqrt{-\ln(X_{\vec{l}})/2},
\end{equation}
where $X_{\vec{l}}$ and $Y_{\vec{l}}$ are random variables uniformly distributed between 0 and 1 for each $\vec{l}$. This is equivalent to generating the Fourier modes of the field as Gaussian random numbers with variance $|u_{\vec{l}}|^2$ as follows:
\begin{equation}
\label{eq:randommodes}
\tilde{\phi}_{\vec{l}}=e^{i 2\pi Y_{\vec{l}}} \sqrt{-\ln(X_{\vec{l}})} \text{ }|u_{\vec{l}}|,
\end{equation}
where $u_{\vec{l}}$ is given by \cref{eq:discmodes} with $a=1$ and $\tau=0$. From here, we first apply the iDFT \cref{eq:iDFT} and then add the background value of the inflaton to obtain the initial field configuration on the lattice. The fluctuations of the time derivative of the scalar field $\partial_\tau\tilde{\phi}_{\vec{l}}$ are generated in the same way using the time derivative of the mode functions $\partial_\tau{u}_{\vec{l}}$ and using the same realizations of $X_{\vec{l}}$ and $Y_{\vec{l}}$. Note that we do not adopt the same procedure of LATTICEEASY for generating the initial field configuration, which is known to have a bug, as first noticed in \cite{Frolov_2008}. However the Fourier transforms in our code are computed through the same routines of LATTICEEASY, contained in the script FFTEASY that is publicly available online~\cite{ffteasy}. 
\section{Outputs of the code}
\label{app:output}
We now summarize the various outputs of the code. 
The background quantities are computed as averages over the box. We output quantities such as the average of the field $\bar{\phi}$, its derivative $\bar{\phi}^\prime$, and the energy density and pressure of the field $\rho_{\rm lat}$ and $p_{\rm lat}$.

The code will also output the power spectrum of field fluctuations $P_{\phi}(\vec{k})$, which in all relevant applications will depend only on the absolute value of the momentum $P_{\phi}(k)$. In order to do so, we first take the DFT to obtain $|\tilde{\phi}_{l_1,l_2,l_3}|^2$. Then, after normalizing by a factor $L^{-3}$ to get the physical power spectrum of the mode functions (see the discussion in \cref{app:icf}), we average over spherical bins to obtain the one dimensional isotropic power spectrum $P_\ell$. This is done averaging $|\tilde{\phi}_{l_1,l_2,l_3}|^2$ over all lattice points such that $\text{int}(|\vec{l}|)=\ell$, where $\ell$ is the bin number. Then, a comoving momentum $k_{\rm lat,\ell}$ is associated to each bin by averaging the absolute value of \cref{eq:modes} over the bin. Note that the procedure for associating the momentum to each bin is different from the one of LATTICEEASY, where the momenta associated to the bins are simply $2\pi\ell/L$. This leads to a distortion in the output momenta of LATTICEEASY, which is independent of $N$ and can lead to a difference of up to $20\%$ in the IR\footnote{Note that this distortion can also be relevant in generating the initial conditions.}. We output the power spectrum for modes only up to the Nyquist frequency $k_{\rm Nyquist}=2\pi \sqrt{3} N_{\rm Nyquist}/L$, where $N_{\rm Nyquist}=N/2$,  because they contain all the physical information.

 The dimensionless power spectrum $\mathcal{P}^{\rm( lat)}_{\phi}(k)$ is obtained multiplying $P_\ell$ by $k_{\rm lat,\ell}^2/(2\pi^2)$ (left plots of \cref{fig:finalPS,fig:finalPS_step}). However, as we discuss in \cref{sec:resultsslow}, we can successfully reproduce the results of the continuous theory at all scales only if we multiply the dimension-full power spectrum $P(k_l)$ by $k_{\rm eff,\ell}^2/(2\pi^2)$ instead of $k_{\rm lat, \ell}^2/(2\pi^2)$, where $k_{\rm eff}$ is obtained averaging \cref{eq:keff} over the same spherical bins (right plots of \cref{fig:finalPS,fig:finalPS_step}).
\subsection{Energy conservation}
\label{app:energy}
\begin{figure}
	\centering
	
	\begin{tikzpicture}
	\node (img) {\includegraphics[width=6cm]{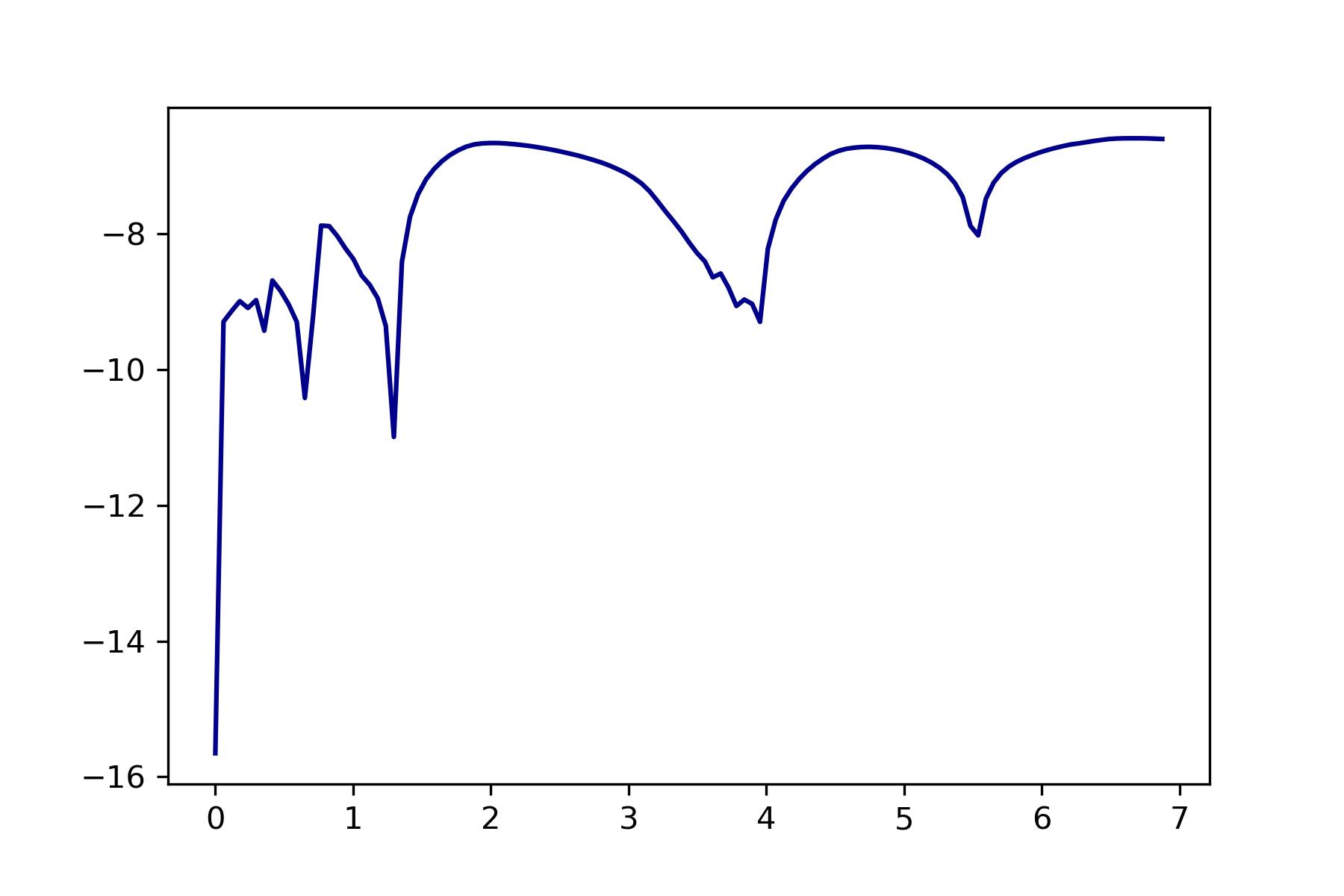}};
	
	\node [rotate=0,text width=0.01cm,align=center] at (-5.,0){ $\log_{10}{|E-1|}$};
	\node [text width=0.01cm,align=center] at (0,-2.2){$N_e$};

	\node (img2) at (6.5,0) {\includegraphics[width=6cm]{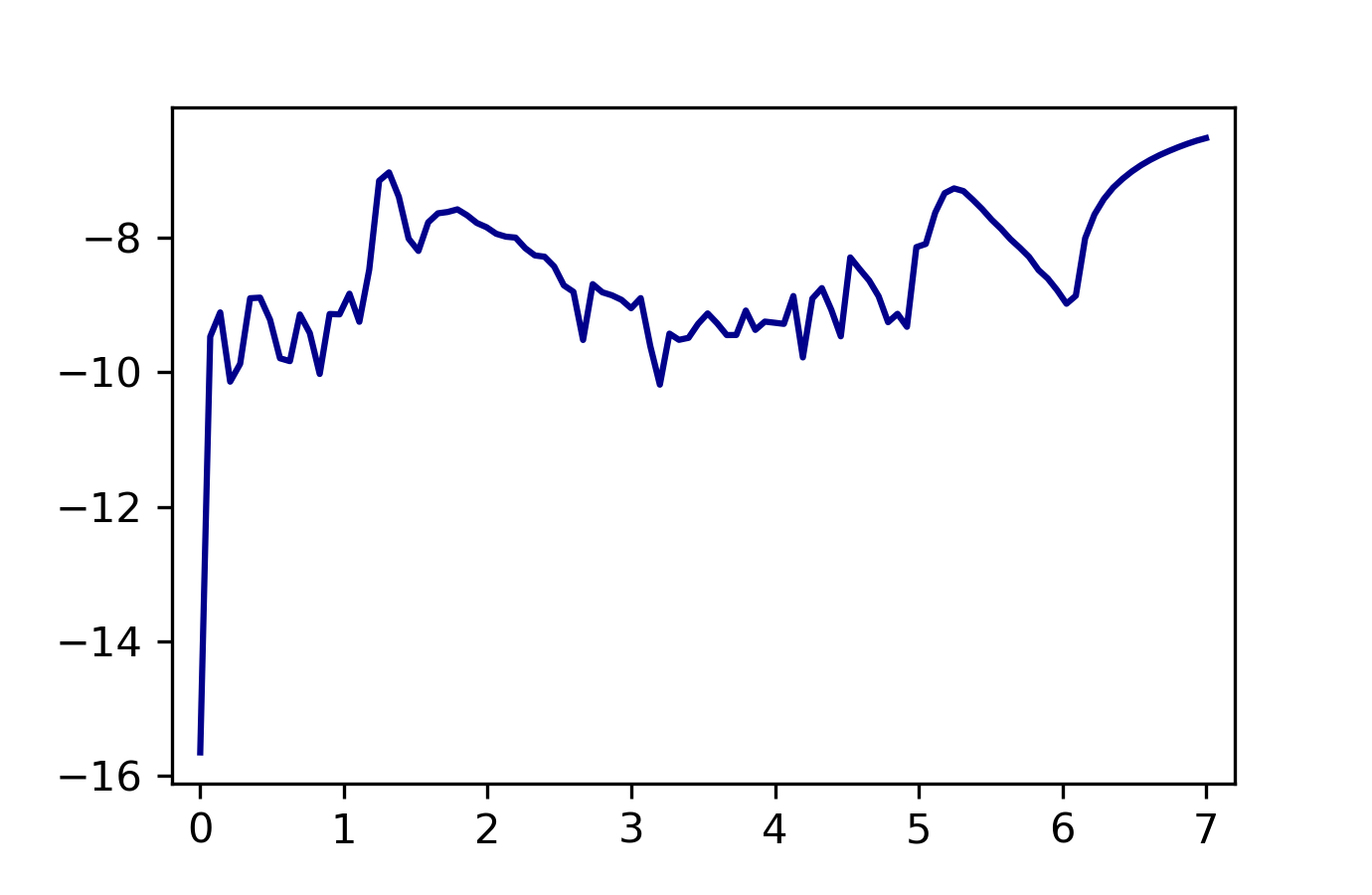}} ;
	
	\node [text width=0.01cm,align=center] at (+6.5,-2.2){$N_e$};

	\end{tikzpicture}

	\caption{Plot of energy conservation during the simulation. On the left we show the result for the standard slow roll potential of \cref{sec:resultsslow}, while on the right we show the result for the step potential of \cref{sec:resultsstep} with $s=0.01$ and $d=0.005$.}
	\label{fig:energy}
\end{figure}
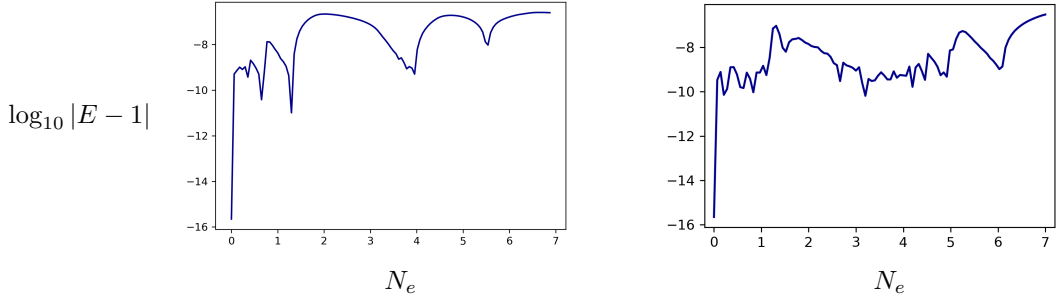

 In order to check energy conservation during the evolution of the system, we define the following quantity:
\begin{equation}
\label{eq:cons}
E = \frac{3\mathcal{H}^2}{\rho a^2}=\frac{3R^2a^{2r}{a^\prime}^2}{a^4 {\rho}_{\rm lat}}.
\end{equation}
We use this quantity, which should be close to 1, to quantify energy conservation in our code. In \cref{fig:energy} we show the plots of energy conservation in the two examples of \cref{sec:results}. From these plots we see that energy is conserved at $10^{-7}$ level in both cases, and the same holds for all the examples discussed in this paper.

\section{Different stencils for the Laplacian operator}
In this section we consider different stencils for the Laplacian. We refer to the Laplacian considered in the main text of \cref{eq:discretelaplacian} and its corresponding effective momentum as $L^{(2)}[\phi]_{i_1,i_2,i_3}$ and $k^{(2)}_{\rm eff}$, where the $2$ refers to the second order of the stencil.
The first one we consider is the following 4th order stencil, which has a similar structure of $L^{(2)}[\phi]$ but involves more points:
\begin{equation}
	L^{(4)}[\phi]_{i_1,i_2,i_3}=\frac{1}{dx^2}\sum_{a_1,a_2,a_3}c_{a_1,a_2,a_3}\phi_{i_1+a_1,i_2+a_2,i_3+a_3},
\end{equation}
where the only non-zero coefficients are $c_{\pm1,0,0}=c_{0,\pm1,0}=c_{0,0,\pm1}=4/3$, $c_{\pm2,0,0}=c_{0,\pm2,0}=c_{0,0,\pm2}=-1/12$ and $c_{0,0,0}=-15/2$.

\begin{figure}
	\centering
	
	\begin{tikzpicture}
	\node (img) {\includegraphics[width=12cm]{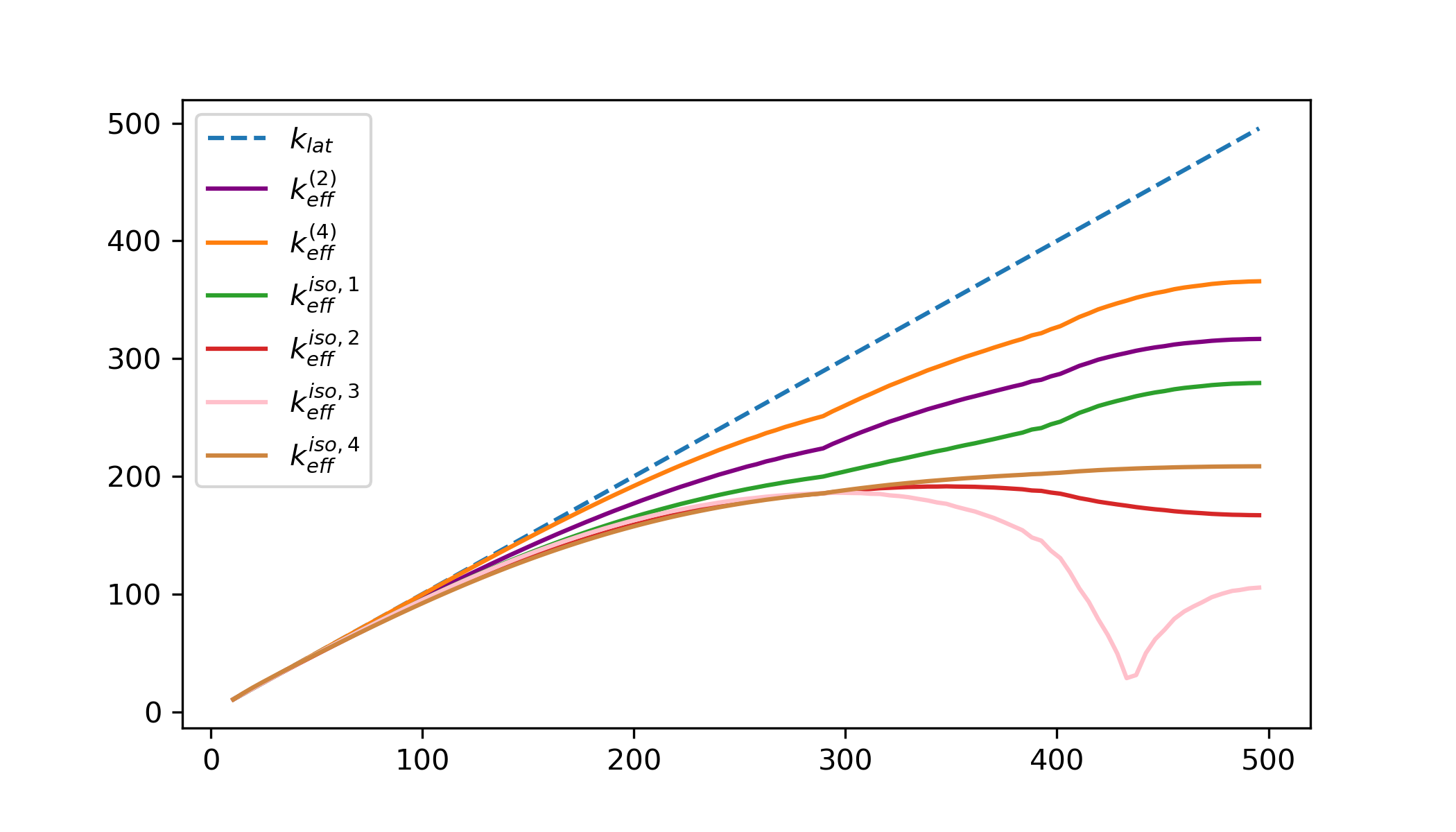}};
\node [text width=0.01cm,align=center] at (-6.5,0){$k_{\rm eff}/m$};
	\node [text width=0.01cm,align=center] at (-.5,-3.5){$k_{\rm lat}/m$};

	\end{tikzpicture}

	\caption{Comparison of the effective momenta coming from different Laplacian operators defined in \cref{app:diff} for a lattice with $N=128$ and $L=1.4/m$.}
	\label{fig:compare}
	
\end{figure}

Next, we consider the anisotropic second order stencils defined in \cite{stencils}. We display the coefficients associated to these stencil as:
\begin{equation}
\begin{bmatrix}
c_{1,1,1}      & c_{0,1,1} & c_{-1,1,1}  \\
c_{1,0,1}      & c_{0,0,1} & c_{-1,0,1}   \\
c_{1,-1,1}       & c_{0,-1,1}   & c_{-1,-1,1}  
\end{bmatrix}
\begin{bmatrix}
c_{1,1,0}      & c_{0,1,0} & c_{-1,1,0}  \\
c_{1,0,0}      & c_{0,0,0} & c_{-1,0,0}   \\
c_{1,-1,0}       & c_{1,-1,0}   & c_{-1,-1,0}  
\end{bmatrix}
\begin{bmatrix}
c_{1,1,-1}      & c_{0,1,-1} & c_{-1,1,-1}  \\
c_{1,0,-1}      & c_{0,0,-1} & c_{-1,0,-1}   \\
c_{1,-1,-1}       & c_{1,-1,-1}   & c_{-1,-1,-1}  
\end{bmatrix}.
\end{equation}
With this convention, we can display the 4 anisotropic stencils of \cite{stencils} as:
\begin{equation}
\quad \quad \quad\,\, L^{\rm iso,1}[\phi]:\quad \begin{bmatrix}
1/12     & 0& 1/12   \\
0      & 2/3 & 0   \\
1/12     & 0  & 1/12 
\end{bmatrix}
\begin{bmatrix}
0     &  \frac{2}{3} & 0 \\
 2/3   & -14/3 & 2/3  \\
0      &  \frac{2}{3}   &0
\end{bmatrix}
\begin{bmatrix}
1/12    & 0& 1/12    \\
0      & 2/3 & 0   \\
1/12     & 0  &1/12   
\end{bmatrix}
\quad\quad\quad\,\,\,\,\,\,
\end{equation}
\begin{equation}
L^{\rm iso,2}[\phi]:\quad \begin{bmatrix}
0   & 1/6& 0  \\
1/6     & 1/3 &  1/6   \\
0      &  1/6  & 0
\end{bmatrix}
\begin{bmatrix}
 1/6     &  1/3 &  1/6 \\
1/3    & -4 &  1/3  \\
1/6     &  1/3   & 1/6
\end{bmatrix}
\begin{bmatrix}
0   &  1/6 &0   \\
1/6    & 1/3 &  1/6  \\
0      &  1/6  & 0  
\end{bmatrix}
\quad\quad\quad\quad\quad\quad\,
\end{equation}
\begin{equation}
\,L^{\rm iso,3}[\phi]:\quad \begin{bmatrix}
-1/12   & 1/3& -1/12  \\
1/3     & 0 &  1/3   \\
-1/12      &  1/3  & -1/12
\end{bmatrix}
\begin{bmatrix}
1/3     &  0 &  1/3 \\
0   & -10/3 &  0  \\
1/3     &  0   & 1/3
\end{bmatrix}
\begin{bmatrix}
-1/12   &  1/3 &-1/12   \\
1/3    & 0 &  1/3  \\
-1/12     &  1/3  & -1/12  
\end{bmatrix}
\end{equation}
\begin{equation}
L^{\rm iso,4}[\phi]:\quad \begin{bmatrix}
1/30   & 1/10& 1/30  \\
1/10    & 7/15 &  1/10   \\
1/30      &  1/10  & 1/30
\end{bmatrix}
\begin{bmatrix}
1/10     &  7/15 &  1/10\\
7/15   & -64/15 &  7/15  \\
1/10     &  7/15   & 1/10
\end{bmatrix}
\begin{bmatrix}
1/30   &  1/10 &1/30   \\
1/3    & 7/15 &  1/3  \\
1/30     &  1/10  & 1/30  
\end{bmatrix}.
\end{equation}
For each stencil $L^{i}$ we refer to its corresponding effective momentum as $k_{\rm eff}^{i}$. We avoid writing the lengthy expressions for all the effective momenta, but we plot them in \cref{fig:compare} for a lattice with $N=128$ and $L=1.4/m$.
All the $k^{\rm iso,i}_{\rm eff}$ are real, with the exception of $k^{\rm iso, 3}_{\rm eff}$ which becomes purely imaginary around $k_{\rm lat}\simeq435m$ (we show the absolute value of $k^{\rm iso, 3}_{\rm eff}$ in the plot). From this plot we can see that only $L^{(4)}$ performs better than $L^{(2)}$ in terms of $k_{\rm eff,max}$ and in terms of overall deviation from $k_{\rm lat}$, while the other isotropic stencils are significantly worse in this sense. The isotropic stencils, however, might perform better under other points of view. For example, these stencils do not have directional dependence in the second order truncation term in real space \cite{stencils}, contrarily to $L^{(2)}$ and $L^{(4)}$.

\begin{figure}
	\centering
	
	\begin{tikzpicture}
	\node (img) {\includegraphics[width=8.cm]{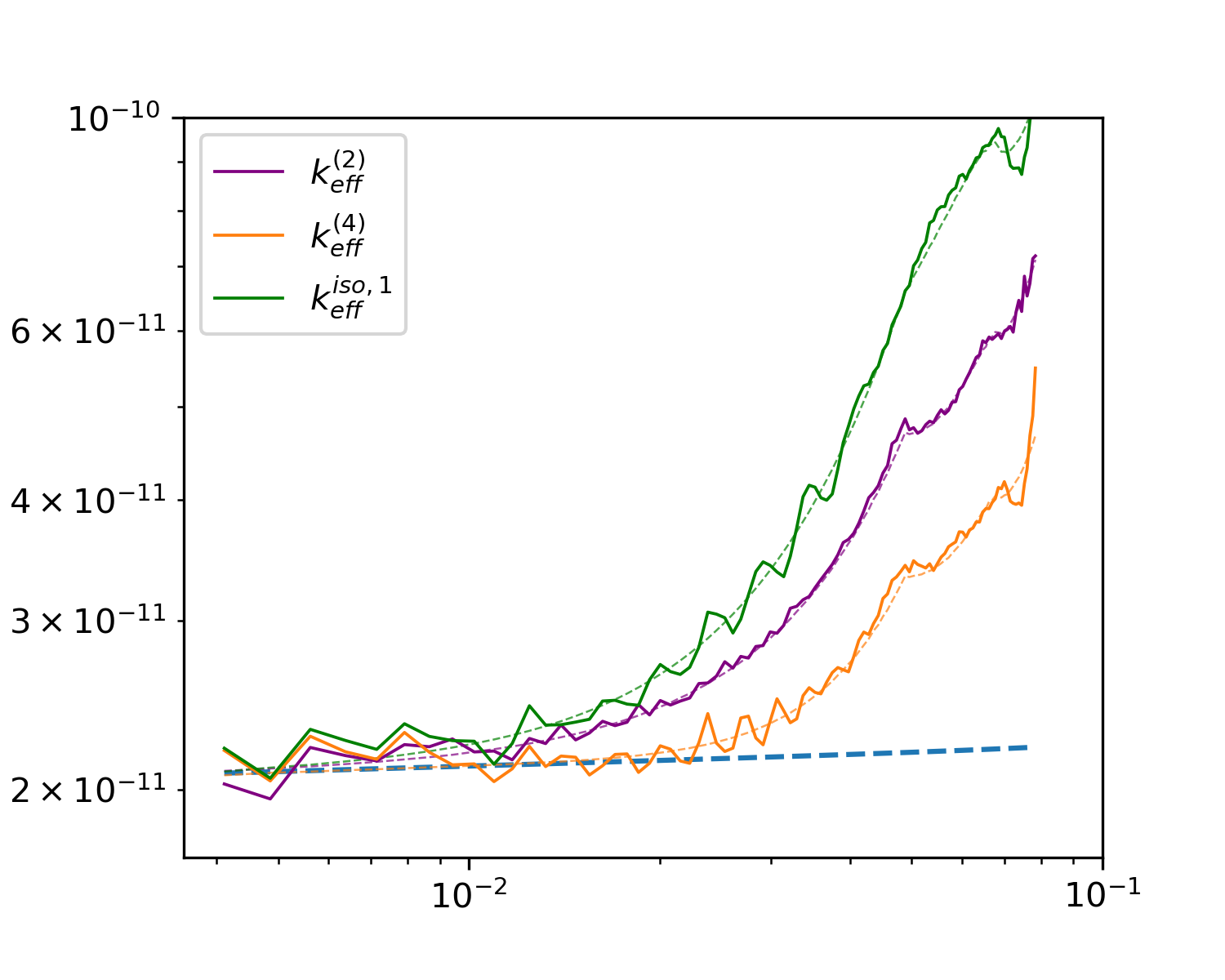}};
	
	\node [rotate=0,text width=0.01cm,align=center] at (-4.6,0){ $\mathcal{P}_{\phi}$};
	\node [text width=0.01cm,align=center] at (0,-3){$k_{\rm lat}/aH$};

	\node (img2) at (7.5,0) {\includegraphics[width=8.cm]{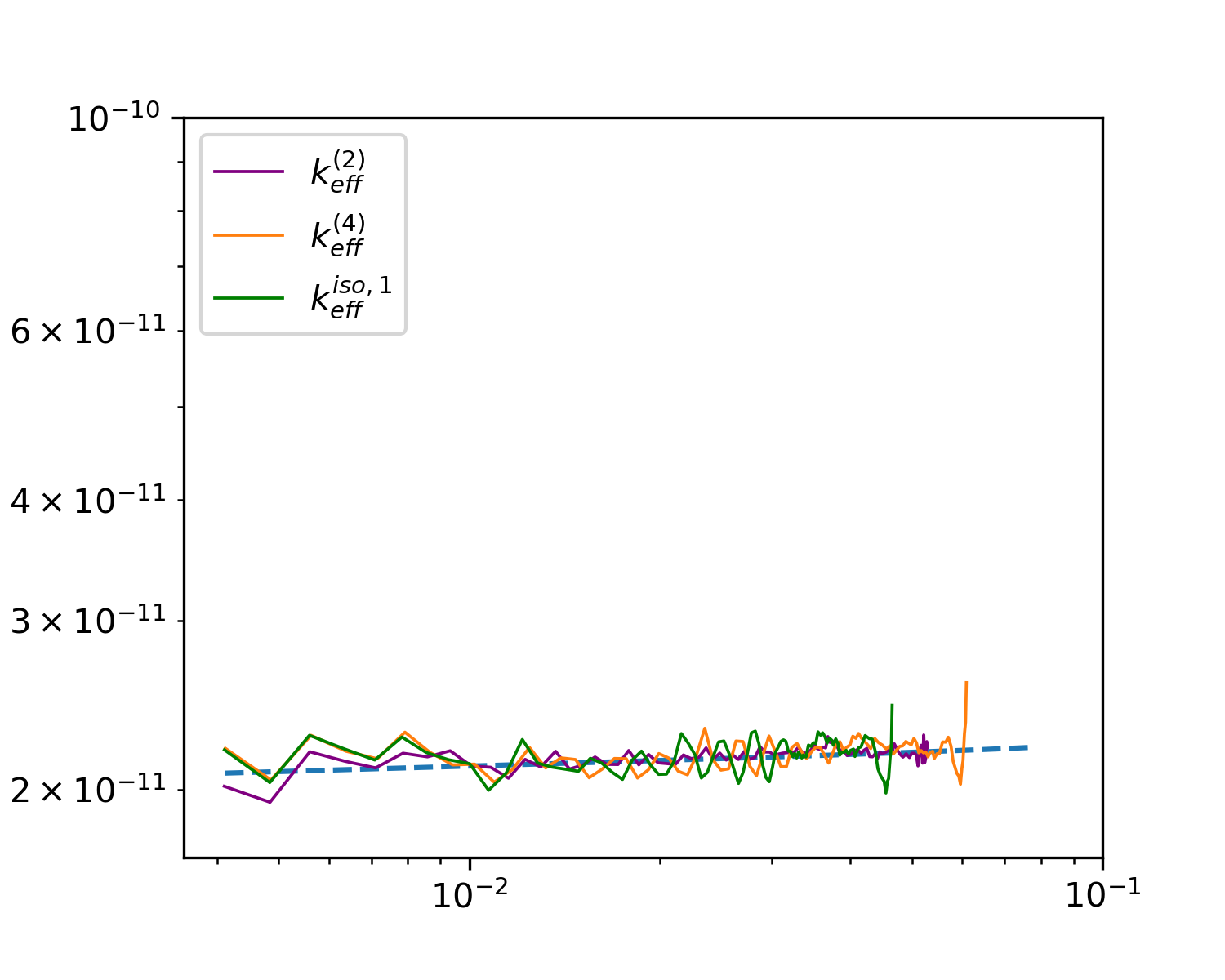}} ;
	
	\node [text width=0.01cm,align=center] at (0+7,-3){$k_{\text{eff}}/aH$};

	\end{tikzpicture}

	\caption{Plot of the final power spectrum computed from the lattice simulation for different stencils for the Laplacian operator. On the left panel we show the lattice results without taking into account the modified dispersion relation and we compare it to the continuous result (blue dashed line). The dashed lines in the left plot are the predictions for discrete dynamics computed from \cref{eq:discthprediction}. On the right, we show the results after the identification $k_{\rm eff}\leftrightarrow k$.}
	\label{fig:finalPS_stencils}
\end{figure}
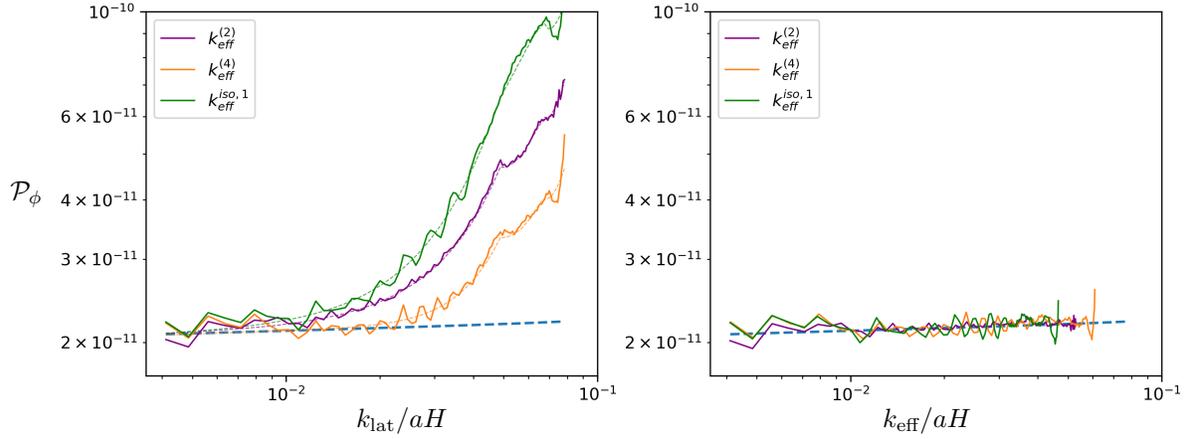

In \cref{fig:finalPS_stencils} we show the final power spectrum computed from simulations with different stencils for the Laplacian. We run these simulations with the $\frac{1}{2}m^2\phi^2$ model and with the same parameters of \cref{sec:resultsslow}. We compare results from $L^{(2)}$, $L^{(4)}$ and the isotropic stencil $L^{\rm iso,1}$. This figure is analogous to \cref{fig:finalPS}, and the dashed lines in the left plot are the analytical predictions for discrete dynamics computed from \cref{eq:discthprediction}. In all these cases, we can see that the identification $k_{\rm eff}\leftrightarrow k$ allows us to recover the continuous result (right plot of \cref{fig:finalPS_stencils}). The same result can be obtained with the other isotropic stencils $L^{iso,i}$, that we do not show in order to make the plots more readable. For $L^{\rm iso, 2}$ and $L^{\rm iso, 3}$, however, this is true only up to a certain momentum cutoff after which $k_{\rm eff}(k_{\rm lat})$ starts decreasing, making the modes unphysical (see \cref{fig:compare}).

\bibliographystyle{jhep}
\bibliography{InflationSim}

\label{app:diff}
\end{document}